\documentclass[showpacs,preprintnumbers,amssymb,reprint,onecolumn,notitlepage,aps,superscriptaddress,showkeys,amsmath,floatfix]{revtex4-2}

\usepackage[utf8]{inputenc}
\usepackage{graphicx}
\usepackage{dcolumn}
\usepackage[dvipsnames]{xcolor}
\usepackage{bm}
\usepackage{amsmath}
\usepackage{amssymb}
\usepackage{epsfig}
\usepackage{amsfonts}
\usepackage{chngcntr} 
\usepackage{lineno,hyperref}
\usepackage{array}
\usepackage{booktabs}
\usepackage{float}
\usepackage{microtype}
\usepackage{multirow}
\usepackage{adjustbox}
\usepackage[english]{babel}
\usepackage{newunicodechar}
\newunicodechar{−}{\ensuremath{-}}
\usepackage{epstopdf}
\usepackage{blindtext}
\usepackage{booktabs}
\usepackage{subcaption}
\usepackage[a4paper, total={6.5in, 10in}]{geometry}
\usepackage{appendix}

\usepackage{etoolbox}
\makeatletter
\patchcmd{\maketitle}{\newpage}{}{}{}
\patchcmd{\maketitle}{\clearpage}{}{}{}
\makeatother
\newcommand{\al}{\alpha}
\newcommand{\beeta}{\beta}
\newcommand{\lam}{\lambda}
\newcommand{\Lam}{\Lambda}

\newcommand{\del}{\delta}

\newcommand{\om}{\omega}
\newcommand{\Om}{\Omega}
\newcommand{\rh}{\rho}
\newcommand{\Del}{\Delta}

\newcommand{\tht}{\theta}

\newcommand{\fs}{f\sigma_8}

\def \be{\begin{equation}}
\def \ee{\end{equation}}
\def \ben{\begin{eqnarray}}
\def \een{\end{eqnarray}}

\def \n{\nonumber}

\def \La{\mathcal{L}}

\begin{document}

\title{Distinguishing Coupled Dark Matter Dark Energy from Kinematic Phantom Crossing}

\author{Samit Ganguly}
\email{samitgphy07@gmail.com}
\affiliation{Department of Physics, University of Calcutta, 92, A.P.C. Road, Kolkata-700009, India}
\affiliation{Department of Physics, Haldia Government College, Haldia, Purba Medinipur 721657, India}

\author{Koushik Dutta}
\email{koushik@iiserkol.ac.in}
\affiliation{Department of Physical Sciences, Indian Institute of Science Education and Research, Kolkata, Mohanpur-741 246, WB, India}

\author{Goutam Manna$^a$}
\email{goutammanna.pkc@gmail.com \\$^a$Corresponding author}
\affiliation{Department of Physics, Prabhat Kumar College, Contai, Purba Medinipur 721404, India}

\date{\today}

\makeatletter
\renewcommand{\frontmatter@abstractwidth}{\dimexpr\textwidth-2cm\relax} 
\makeatother

\begin{abstract}
Recent measurements of cosmic expansion have renewed interest in dark-energy scenarios in which the effective equation of state exhibits phantom-like evolution and may cross $w=-1$. Such behavior, however, does not require a fundamental phantom field: in coupled dark-energy-dark-matter~(CDEDM) models, energy exchange between a canonical scalar field and cold dark matter, happens Lagrangian labels, can reproduce the same effective equation of state evolution as that of phenomenological $w_0w_a$CDM model. We show that this microscopic interaction leaves distinctive, testable signatures in structure growth by forming a two-fluid system in which only cold dark matter feels the scalar-mediated fifth force (and an associated drag term), while baryons remain uncoupled. When combined with neutrino-free streaming, this setup generates a characteristic coupling, neutrino-mass degeneracy, whose leading scaling $\sum m_\nu \propto f_c^2 \alpha^2$ follows analytically from the two-fluid growth equations. Using DESI DR2 BAO, CMB distance information, ACT DR6 lensing, redshift-space distortions, and three supernova compilations, we find that this scaling is recovered for both exponential and inverse-power-law potentials. We further show that the dominant contribution to the growth response arises from the coupling-induced modification of the background evolution, with the fifth force enhancing growth and the drag partially counteracting it. Finally, the interaction induces a running dark-matter mass, changing it at recombination by $4.4\%$--$5.5\%$, so CMB distance calculations must account for the evolving dark-matter density. These correlated growth and recombination effects provide a route to distinguish interaction-driven phantom crossing from purely kinematic dark-energy parametrizations.
\end{abstract}

\keywords{}

\maketitle

\section{Introduction}
Late-time cosmic acceleration was established observationally with Type~Ia supernovae~\cite{Riess1998,Perlmutter1999} and is now tightly constrained by a combination of distance probes, including BAO from large-scale structure surveys~\cite{Eisenstein2005,Cole2005,Alam2017} and the CMB~\cite{Planck2020,actdr6}. These measurements also underpin the ``$H_0$ tension'' between late-Universe distance-ladder determinations and early-Universe inferences~\cite{Abbott2017,GW170817,Verde2019,DiValentino2021,Riess2022}. Recent DESI BAO results have renewed interest in dynamical dark energy and in reconstructions where the dark-energy equation of state evolves and may cross the phantom divide $w=-1$~\cite{Chevallier2001,Linder2003,Copeland2006,DESI2025}. If such phantom-like behaviour were fundamental, it would typically require a genuine phantom degree of freedom and can be accompanied by theoretical difficulties related to null-energy-condition violation and instabilities~\cite{Caldwell2002,Hu2005,Vikman2005}.

A physically well-motivated alternative is that the phantom crossing is only \emph{effective}, because dark matter and dark energy are not separately conserved but interact within a common dark sector. In a coupled dark-energy--dark-matter (CDEDM) model, the individual dark-sector energy--momentum tensors exchange energy and/or momentum while the total remains conserved~\cite{Damour1992,Wetterich1995,Amendola2000,Amendola2004,Simpson2010,Baldi2015,Carrion2024}. This has three immediate consequences that motivate our choice of framework. First, the equation of state inferred from background data is an effective quantity, so a canonical (non-phantom) scalar field can mimic an effective crossing of $w=-1$ without introducing a fundamental phantom sector~\cite{Amendola2000,Amendola2004,Chakraborty2025,Antusch2026}; for a discussion of stability in interacting-fluid descriptions, see Ref.~\cite{Valiviita2008}. Second, because the relative dark-sector densities no longer follow independent dilution laws, the interaction provides a dynamical perspective on the cosmic coincidence problem~\cite{Zlatev1999}. Third, the coupling leaves signatures beyond the background expansion: cold dark matter experiences a scalar-mediated fifth force and a coupling-induced drag term, while baryons remain uncoupled, leading to a genuinely two-fluid growth system~\cite{Peebles1980,Damour1992,MaBertschinger1995,Amendola2000,Amendola2004,Liddle,Dodelson}.

The objective of the present work is to identify observables that can trace the underlying microscopic CDM--scalar interaction even when the background expansion is nearly indistinguishable from a phenomenological $w_0w_a$CDM description~\cite{Chevallier2001,Linder2003}. To this end, we derive a leading-order scaling relation for the coupling--neutrino-mass degeneracy, $\sum m_\nu\propto f_c^2\alpha^2$, which follows from the two-fluid growth equations and from the competition between coupling-enhanced CDM clustering and neutrino free streaming, and we test this prediction against the full posterior~\cite{lesgourgues,BeltranJimenez2026}. We also show that the same interaction induces a running dark-matter mass, so that the cold-dark-matter density at recombination differs from its present value and compressed CMB constraints must be evaluated using recombination-epoch densities throughout. Overall, rather than treating the coupling solely as a way to reproduce a phantom-like effective equation of state, we exploit its correlated signatures in the growth and recombination sectors to distinguish an interaction-driven scenario and thereby constrain its microscopic origin.

The paper is organized as follows. In Sec.~II we introduce the coupled-quintessence framework at the background and linear-perturbation level, including the two-fluid (CDM+baryon) growth system and our numerical evaluation of $f\sigma_8(z)$. Sec.~II also describes the data vector and inference strategy, specifies the likelihood components, and summarizes the priors and nested-sampling settings. In Sec.~III we present the main constraints and their physical implications: we compare the effective EoS with $w_0w_a$CDM, quantify the running dark-matter mass at recombination, perform residual checks, and analyse both the dark-sector energy exchange and the coupling-neutrino-mass degeneracy before commenting on the $H_0$ and $S_8$ tensions. We conclude in Sec.~IV, with additional technical details and robustness tests collected in the appendices.

\section{Theoretical Framework}
We consider a canonical quintessence field $\phi$ coupled to fermionic cold dark matter $\psi_{\rm c}$ through a Yukawa-type interaction~\cite{Wetterich1995,Amendola2000,Amendola2004,Chakraborty2025,Antusch2026},
\ben
\mathcal{L}_{\rm int}\propto -f(\phi/M_{\rm Pl})\,\bar{\psi}_{\rm c}\psi_{\rm c},
\label{1}
\een
where $M_{\rm Pl}$ is the (reduced) Planck mass and $f$ is a dimensionless coupling function. This interaction makes the CDM mass field dependent,
$ m_{\rm c}(\phi)=m_{\rm c}^{(0)}\,\frac{f(\phi)}{f(\phi_0)} $, so that a convenient effective Lagrangian is (see, e.g., Refs.~\cite{Wetterich1995,Amendola2000,Amendola2004})
\ben
\boxed{
\begin{aligned}
\mathcal{L}_{\rm tot}
=&\,\frac{M_{\rm Pl}^{2}}{2}R
-\frac{1}{2}g^{\mu\nu}\partial_{\mu}\phi\,\partial_{\nu}\phi
-V(\phi)
\\[2mm]
&+i\bar{\psi}_{c}\gamma^{\mu}D_{\mu}\psi_{c}
-m_{c}^{(0)}\,\frac{f(\phi)}{f(\phi_{0})}\,\bar{\psi}_{c}\psi_{c}
\\[2mm]
&+i\bar{\psi}_{b}\gamma^{\mu}D_{\mu}\psi_{b}
-m_{b}\bar{\psi}_{b}\psi_{b}
+\mathcal{L}_{\nu}+\mathcal{L}_{\gamma}.
\end{aligned}}
\label{1a}
\een
\noindent
Here $R$ is the Ricci scalar, $g_{\mu\nu}$ is the metric, and $D_{\mu}$ is the covariant derivative. The baryons $\psi_b$ are taken to be uncoupled (constant $m_b$), as required by local equivalence-principle tests, while the interaction allows energy-momentum exchange between $\phi$ and CDM. The remaining terms $\mathcal{L}_\nu$ and $\mathcal{L}_\gamma$ describe standard neutrinos and photons.

The corresponding CDM background density evolves as
\ben
\rho_{\rm c}(a)=\frac{\rho_{\rm c}^{(0)}}{a^3}\,\frac{f(\phi)}{f(\phi_0)}.
\label{2}
\een
The scalar field then evolves according to the modified Klein-Gordon~(KG) equation,
\ben
\ddot{\phi}+3H\dot{\phi}
=-V_{,\phi}
-\frac{\rho_{\rm c}^{(0)}}{a^3}
\frac{f_{,\phi}}{f(\phi_0)},
\label{3}
\een
which can be written in terms of an effective potential,
\ben
V_{\rm eff}(\phi,a)=V(\phi)+
\frac{\rho_{\rm c}^{(0)}}{a^3}
\frac{f(\phi)}{f(\phi_0)}.
\een
Here $V_{,\phi}\equiv dV/d\phi$ and $f_{,\phi}\equiv df/d\phi$. The interaction therefore modifies the scalar dynamics through energy exchange with CDM, while the scalar kinetic term remains canonical.

If the interaction is not explicitly modelled, the background expansion can be recast in terms of an effective dark-energy density~\cite{Amendola2000,Amendola2004},
\ben
\rho_{\rm DE,eff}=
\rho_\phi+
\frac{\rho_{\rm c}^{(0)}}{a^3}
\left[\frac{f(\phi)}{f(\phi_0)}-1\right],
\label{4}
\een
with an associated effective equation of state
\ben
w_{\rm eff}
\equiv\frac{P_\phi}{\rho_{\rm DE,eff}}
=\frac{w_\phi}{1-x},
\qquad
x\equiv
\frac{\rho_{\rm c}^{(0)}}{\rho_\phi a^3}
\left[1-\frac{f(\phi)}{f(\phi_0)}\right].
\label{5}
\een
For a canonical scalar field the kinetic term is positive, $\dot{\phi}^2>0$, and its equation of state satisfies $-1\leq w_\phi\leq 1$. Nevertheless, the 
\emph{effective} equation of state can cross the phantom divide when $1+w_\phi<x<1$~\cite{Amendola2000,Amendola2004}.

The condition $1+w_\phi<x$ requires $f(\phi)<f(\phi_0)$, i.e., energy transfer from CDM to the scalar in this parametrization, while $x<1$ ensures $\rho_{\rm DE, eff}>0$. Phantom behavior is therefore an emergent consequence of dark-sector energy exchange rather than a fundamental negative-kinetic-energy degree of freedom.\\

For our analysis, we adopt the standard exponential coupling~\cite{Wetterich1995, Amendola2000, Amendola2004}
\ben
f(\phi)=\exp\!\left(\frac{\alpha\,\phi}{\sqrt{8\pi}\,M_{\rm Pl}}\right),
\label{7}
\een
for which $\partial_\phi\ln f=\alpha/(\sqrt{8\pi}\,M_{\rm Pl})$ is constant. For $\alpha>0$, the phantom-crossing condition derived above implies $\phi<\phi_0$, which is naturally realised in tracker/scaling quintessence evolution~\cite{Copeland1998,Zlatev1999}.

We consider two theoretically motivated choices for the potential,
\ben
V(\phi)&=V_0\exp\!\left(-\frac{\lambda\,\phi}{\sqrt{8\pi}\,M_{\rm Pl}}\right),
\label{8a}\\
V(\phi)&=V_0\left(\frac{\phi}{M_{\rm Pl}}\right)^{-\lambda},
\label{8}
\een
corresponding respectively to exponential potentials that admit scaling solutions (as in early cosmon/quintessence models) and inverse-power-law tracker potentials~\cite{Wetterich1995,Copeland1998,RatraPeebles1988,Zlatev1999,Copeland2006}. While other well-motivated forms exist (e.g. supergravity-inspired models~\cite{Brax1999}), we restrict attention to these two families (Eqs. \eqref{8a} and \eqref{8}). We constrain $\{\lambda,V_0,\alpha\}$ jointly with the cosmological parameters using the observational datasets described above.\\

All background quantities are integrated using the $e$-fold variable $N\equiv\ln a$, which mitigates stiffness as $a\to0$ and is convenient for our JAX/Diffrax ODE-solver workflow~\cite{JAX2018,Diffrax2020,Kidger2021}. With the exponential coupling $f(\phi)=\exp\!\left(\alpha\phi/\sqrt{8\pi}\,M_{\rm Pl}\right)$ and $f_0\equiv f(\phi_0)$, the dimensionless expansion rate $E\equiv H/H_0$ satisfies
\ben
E^2 = \Om_{c0}\frac{f(\phi)}{f_0}e^{-3N} + \Om_{b0}e^{-3N}
    + \Om_{\gamma0}e^{-4N} + \Om_\nu(N) + \frac{\rh_\phi}{3H_0^2},
\label{A1}
\een
where $\Om_\nu(N)$ is constructed from the sampled $\sum m_\nu$ and $N_{\rm eff}$, with an interpolation through the relativistic-to-non-relativistic transition~\cite{lesgourgues}. Neutrinos are uncoupled to $\phi$. The factor $f(\phi)/f_0$ in the CDM term represents the running CDM mass, not an additional fluid component.\\

Writing $v\equiv\dot\phi$, we rewrite the background evolution equations, Eqs.~\eqref{3} and \eqref{A1}, as a first-order system in $N\equiv\ln a$,
\ben
\frac{d\phi}{dN} &=& \frac{v}{H}, \label{A2}\\
\frac{dv}{dN} &=& \frac{1}{H}\Big[-3Hv - V_{,\phi}
   - 3H_0^2\Om_{c0}e^{-3N}\frac{f_{,\phi}}{f_0}\Big], \label{A3}\\
\frac{dL}{dN} &=& \frac{e^{-N}}{H}, \label{A4}
\een
where $L$ is defined so that the comoving distance is $D_C=c\,L$ (with $c$ in km\,s$^{-1}$)~\cite{Hogg1999}. The last term in Eq.~\eqref{A3} is the coupling source induced by the running CDM mass.

For the exponential potential, $V_{,\phi}=-(\lambda/\sqrt{8\pi})V$ and $V_{,\phi\phi}=(\lambda/\sqrt{8\pi})^2V$, while for the inverse power-law potential $V_{,\phi}=-\lambda V/\phi$ and $V_{,\phi\phi}=\lambda(\lambda+1)V/\phi^2$. For the exponential coupling of Eq.~\eqref{7}, $f_{,\phi}=(\alpha/\sqrt{8\pi}M_{\rm Pl})f$.

Note that $\lambda$ has different meanings in the two potentials: it is a dimensionless slope parameter in the exponential case, but a power-law index in the inverse-power-law case. Correspondingly, the tracker parameter $\Gamma\equiv V V_{,\phi\phi}/V_{,\phi}^2$ equals $1$ for the exponential potential and $1+1/\lambda$ for the inverse power law, leading to different posterior ranges (Table~\ref{TableI}).

Radiation is fixed by the CMB temperature~\cite{Fixsen2009} and $N_{\rm eff}$, with $\Om_{b0}=\om_b/h^2$ and $\Om_{c0}=\Om_{m0}-\Om_{b0}-\Om_{\nu0}$. We therefore sample seven parameters,
$\tht=\{\Om_{m0},H_0,\om_b,N_{\rm eff},\sum m_\nu,\lambda,\alpha\}$.\\

The dark-sector coupling mediates an additional long-range interaction between dark-matter particles. In the quasi-static, sub-Hubble limit ($k\gg aH$), where time derivatives of the metric and scalar perturbations are subdominant to spatial gradients, the linearised system reduces to a modified Poisson equation with an effective gravitational coupling~\cite{MaBertschinger1995,Amendola2000,Dodelson,Damour1992},
\ben
\frac{G_{\rm eff}}{G_N}
=1+\frac{2\beeta^2}{1+a^2V_{,\phi\phi}/k^2},
\label{9}
\een
where the denominator accounts for the finite Compton wavelength of the scalar. Since baryons are assumed to be uncoupled, CDM and baryons obey distinct growth equations~\cite{Amendola2000,MaBertschinger1995},
\ben
&&\del_c''+
\left(2+\frac{d\ln E}{dN}-\beeta\phi'\right)\del_c'
 =\frac{3}{2E^2}
\left[\Om_c\frac{G_{\rm eff}}{G_N}\del_c+\Om_b\del_b\right],\n \\
&&\del_b''+
\left(2+\frac{d\ln E}{dN}\right)\del_b'
 =\frac{3}{2E^2}
\left[\Om_c\del_c+\Om_b\del_b\right],
\label{10}
\een
with $N\equiv\ln a$, $\Om_c=\Om_{c0}(f/f_0)e^{-3N}$, and $\Om_b=\Om_{b0}e^{-3N}$. The total matter perturbation entering the observables is $\del_m=(\Om_{c0}\del_c+\Om_{b0}\del_b)/(\Om_{c0}+\Om_{b0})$.

Equation~\eqref{10} highlights a key difference from $w_0w_a$CDM: the coupling affects growth through (i) an extra friction term $-\beeta\phi'$ in the CDM equation, (ii) a modified gravitational source $G_{\rm eff}$, and (iii) the time dependence of the CDM background density through $f(\phi)$. These effects generically imply $\del_c/\del_b\neq1$, and can therefore modify the growth history even when the background expansion is nearly degenerate with $w_0w_a$CDM~\cite{Linder2005}.

In our numerical implementation, we therefore evolve the CDM and baryon perturbations separately. The combined background+growth system, consisting of Eqs.~\eqref{A2}--\eqref{A4} and \eqref{10}, is integrated as the seven-component state vector
\ben
\mathbf{Y}=[\phi,\;v,\;L,\;\del_c,\;\del_c',\;\del_b,\;\del_b'].
\label{A7}
\een

However, the initial conditions for the two species are different. Deep in matter domination the coupled CDM perturbation grows with a modified power-law index~\cite{Amendola2000,Dodelson},
\ben
p_g=\tfrac14\big(-1+\sqrt{1+24(1+2\beta^2)}\big),
\qquad \beta\equiv\frac{\al}{\sqrt{8\pi}},
\label{A8}
\een
so we set $\del_c(z_{\rm ini})=a_{\rm ini}^{p_g}$ and $\del_c'(z_{\rm ini})=p_g\,a_{\rm ini}^{p_g}$. Here $1+2\beta^2$ is the \emph{unscreened} value of $G_{\rm eff}/G_N$, since our initial epoch is chosen to lie well inside the scalar Compton radius. Baryons are uncoupled and therefore follow the standard growing mode, $\del_b(z_{\rm ini})=\del_b'(z_{\rm ini})=a_{\rm ini}$~\cite{Peebles1980,MaBertschinger1995}.

The total matter perturbation that enters the observables is defined as the present-day-density-weighted combination
\ben
\del_m=\frac{\Om_{c0}\del_c+\Om_{b0}\del_b}{\Om_{c0}+\Om_{b0}}.
\label{A9}
\een

We solve the growth system on Gauss-Legendre nodes in wavenumber $k$ using \texttt{jax.vmap}~\cite{JAX2018}. From the linear matter power spectrum $P(k,z)$ we compute
\ben
\sigma_8^2(z)=\frac{1}{2\pi^2}\int k^2P(k,z)\,W^2(kR)\,dk,
\qquad R=8\,h^{-1}{\rm Mpc},
\label{A10}
\een
where $W(x)=3(\sin x-x\cos x)/x^3$ is the spherical top-hat window function~\cite{Peebles1980,Dodelson}. We then obtain the redshift-space observable $f\sigma_8(z)$ by differentiating $\sigma_8(z)$ with respect to $\ln a$, i.e. $f\sigma_8=d\sigma_8/d\ln a$~\cite{Peebles1980,Dodelson}. The linear power spectrum is evaluated with CosmoPower-JAX~\cite{cosmopowerjax}, trained on CAMB spectra~\cite{CAMB2000}.

\section{Data and inference}
We constrain our models using Type~Ia supernova distances from DES-SN5YR~\cite{DES2024}, Pantheon$+$~\cite{Brout2022} (building on earlier compilations such as JLA~\cite{Betoule2014} and Pantheon~\cite{Scolnic2018}), and Union3~\cite{Rubin2023}; BAO measurements from DESI~DR2~\cite{DESI2025,DESILya} (see also the general BAO overview in Ref.~\cite{Aubourg2015}). The compressed \textit{Planck}~2018 CMB distance prior~\cite{Zhai2020}, obtained from \textit{Planck}~2018 likelihood chains~\cite{Aghanim2018} using the distance-prior methodology of Refs.~\cite{Chen2019,Zhai2020}, ACT~DR6 CMB lensing~\cite{actdr6}, and redshift-space distortion (RSD) growth data~\cite{Blake2012,NPP2017}. We do not include multi-probe weak-lensing+clustering likelihoods (e.g.\ KiDS-1000 and DES-Y3 3$\times$2pt analyses~\cite{KiDS2x3pt,DES3x2pt}) to keep the inference focused on the growth-sector information carried by $f\sigma_8$. The supernova absolute magnitude is analytically marginalized~\cite{Conley2011}, the full BAO covariance is retained, and the CMB likelihood uses the five-observable compression $(R,\ell_A,\om_b,\om_c,N_{\rm eff})$~\cite{Zhai2020}, with the sound horizon evaluated by direct integration. Because the dark-matter mass runs in CDEDM, the recombination-epoch cold-dark-matter density parameter exceeds its present value by $4.4\%$--$5.5\%$. We therefore evaluate the compressed-CMB observables and both sound horizons using recombination-epoch densities throughout. Finally, we propagate the correlated uncertainty associated with the primordial fluctuation amplitude as an additional rank-one contribution to the covariance.

We sample the summed neutrino mass, $\sum m_\nu$, jointly with the CDEDM parameters so that the degeneracy with the coupling strength $\alpha$ can be tested directly. For a fair comparison, we adopt the same priors, likelihood, and sampling strategy for both CDEDM and $w_0w_a$CDM, and we compute posterior distributions and Bayesian evidences using \texttt{dynesty}~\cite{dynesty2020}.

\subsection{Likelihood}

All model-dataset combinations are analyzed with the same likelihood implementation, prior set, and sampler; any differences in the posteriors therefore reflect the model choice.

\textbf{Supernovae:} We include one compilation at a time—DES-SN5YR ($1820$), Pantheon$+$ with $z\ge0.01$ ($1588$), or Union3 ($22$), and analytically marginalize over the absolute magnitude~\cite{DES2024,Brout2022,Rubin2023,Conley2011}.

\textbf{BAO:} We use the DESI DR2 BAO data vector as the \textit{thirteen} independent quantities, with the full inverse covariance~\cite{DESI2025,DESILya}. For $z\ge0.51$ the tabulated $D_V$ is constructed from $(D_M,D_H)$; treating all nineteen tabulated numbers as independent, which would therefore double-count information~\cite{DESI2025,DESILya}.

\textbf{CMB:} We adopt the five-observable distance-prior compression $(R,\ell_A,\om_b,\om_c,N_{\rm eff})$, derived from a \textit{Planck} chain in which both $N_{\rm eff}$ and $\sum m_\nu$ vary~\cite{Zhai2020}. The more common three-element compression assumes fixed neutrino parameters and is not applicable here. We compute the sound horizon by direct integration of $c_s/H$ back to $z=10^6$, rather than using a fitting formula~\cite{Zhai2020}.

\textbf{CMB lensing:} We include the ACT DR6 constraint $\sigma_8\Om_m^{0.25}=0.599\pm0.017$~\cite{actdr6} as a likelihood term.

\textbf{Growth:} We use the eighteen-point Gold-2017 $\fs$ compilation~\cite{NPP2017,Blake2012} rather than the larger 63-point set, because $38$ of the $63$ points lie at repeated redshifts and are largely re-analyses of the same galaxies; with a diagonal covariance they would be treated as independent and would spuriously tighten constraints on $\al$, which is primarily driven by $\fs$. The WiggleZ points at $z=0.44$, $0.60$, and $0.73$ are included with their measured $3\times3$ covariance~\cite{Blake2012}. Each point is corrected for the Alcock-Paczynski effect via $q(z)=H D_A/(H^{\rm fid}D_A^{\rm fid})$ evaluation using that measurement's fiducial cosmology (here $\Om_m^{\rm fid}=0.25$--$0.31$ across surveys) and applied multiplicatively to the theory prediction. The correction is at the $\sim1\%$ level and is therefore not optional. The observable is defined in the linear Kaiser limit~\cite{Kaiser1987} and uses standard cosmological distance conventions~\cite{Hogg1999}.

We note that the BOSS/eBOSS RSD volumes partly overlap the DESI DR2 BAO sample, so the combined constraint may be mildly optimistic.

Because $A_s$ and $n_s$ carry no information in the compressed CMB likelihood and are held fixed, we propagate the associated fractional amplitude uncertainty, $\sigma_{\rm frac}=0.007955$, as a rank-one fully correlated contribution to both the lensing and RSD covariances,
\ben
\mathsf{C}_{\rm tot}=\mathsf{C}_{\rm RSD}+\sigma_{\rm frac}^2\,\mathbf{p}\mathbf{p}^{T},
\label{A11}
\een
where $\mathbf{p}$ is the predicted $\fs$. A shared amplitude shifts all points coherently, so treating the uncertainty as diagonal would incorrectly allow it to average down. Writing $\mathsf{C}\equiv\mathsf{C}_{\rm RSD}$, Sherman--Morrison gives
\ben
\mathbf{r}^T\mathsf{C}_{\rm tot}^{-1}\mathbf{r}
 = \mathbf{r}^T\mathsf{C}^{-1}\mathbf{r}
 - \frac{\sigma_{\rm frac}^2(\mathbf{p}^T\mathsf{C}^{-1}\mathbf{r})^2}
        {1+\sigma_{\rm frac}^2\,\mathbf{p}^T\mathsf{C}^{-1}\mathbf{p}},
\label{A12}
\een
so no matrix inversion is required per likelihood call. Since $\mathsf{C}_{\rm tot}$ depends on parameters through $\mathbf{p}$, we retain the log-determinant term; by the matrix determinant lemma it is $\ln\,(1+\sigma_{\rm frac}^2\mathbf{p}^T\mathsf{C}^{-1}\mathbf{p}\bigr)$.

\subsection{Priors and sampling}\label{sec:priors}

We adopt uniform priors for all parameters: $\Om_{m0}\in[0.20,0.40]$, $H_0\in[60,80]$, $\om_b\in[0.010,0.030]$, $N_{\rm eff}\in[2.00,4.50]$, $\sum m_\nu\in[0.06,0.60]\,$eV, $\al\in[0,1]$, and $\lam\in[0.01,9.00]$ for the exponential potential or $\lam\in[0.01,1.50]$ for the inverse power-law potential.

The lower prior bound on $\sum m_\nu$ is set by the minimum allowed by neutrino oscillation experiments, i.e. an external terrestrial constraint rather than a cosmological assumption.

We impose a one-sided prior on the coupling because realizing $w_{\rm eff}<-1$ requires $\al>0$. Since the null value ($\al=0$) lies on the prior boundary, a symmetric ``$N\sigma$'' detection significance is not well defined; we therefore report one-sided constraints on $\al$. The upper bound is motivated by fifth-force considerations, as for $G_{\rm eff}/G_N = 1+\al^2/(4\pi)$: $\al=1$ implies an $\simeq 8\%$ enhancement, whereas $\al=10$ would yield nearly an order-of-magnitude increase, which is incompatible with structure formation.

Posterior samples and Bayesian evidences are obtained with dynamic nested sampling using \texttt{dynesty}~\cite{dynesty2020}, with \texttt{bound=`multi'}, \texttt{sample=`rwalk'}, $n_{\rm live}=1000$, $n_{\rm batch}=500$, $30$ walks, a maximum of six batches, and a target of $5000$ effective posterior samples. These settings are used for all runs.

\section{Results}

In this section, we present the results of our MCMC analysis across the different supernova compilations and scalar-field potentials. We organize the discussion into subsections~A--G where best-fit values and posteriors, a comparison of the effective EoS, the running dark-matter mass at recombination, a residual check, dark-sector energy exchange, the coupling--neutrino-mass degeneracy, and implications for cosmological tensions have been discussed.

\subsection{Best-fit values and posteriors}

\begin{figure}
\centering
\includegraphics[width=0.49\textwidth]{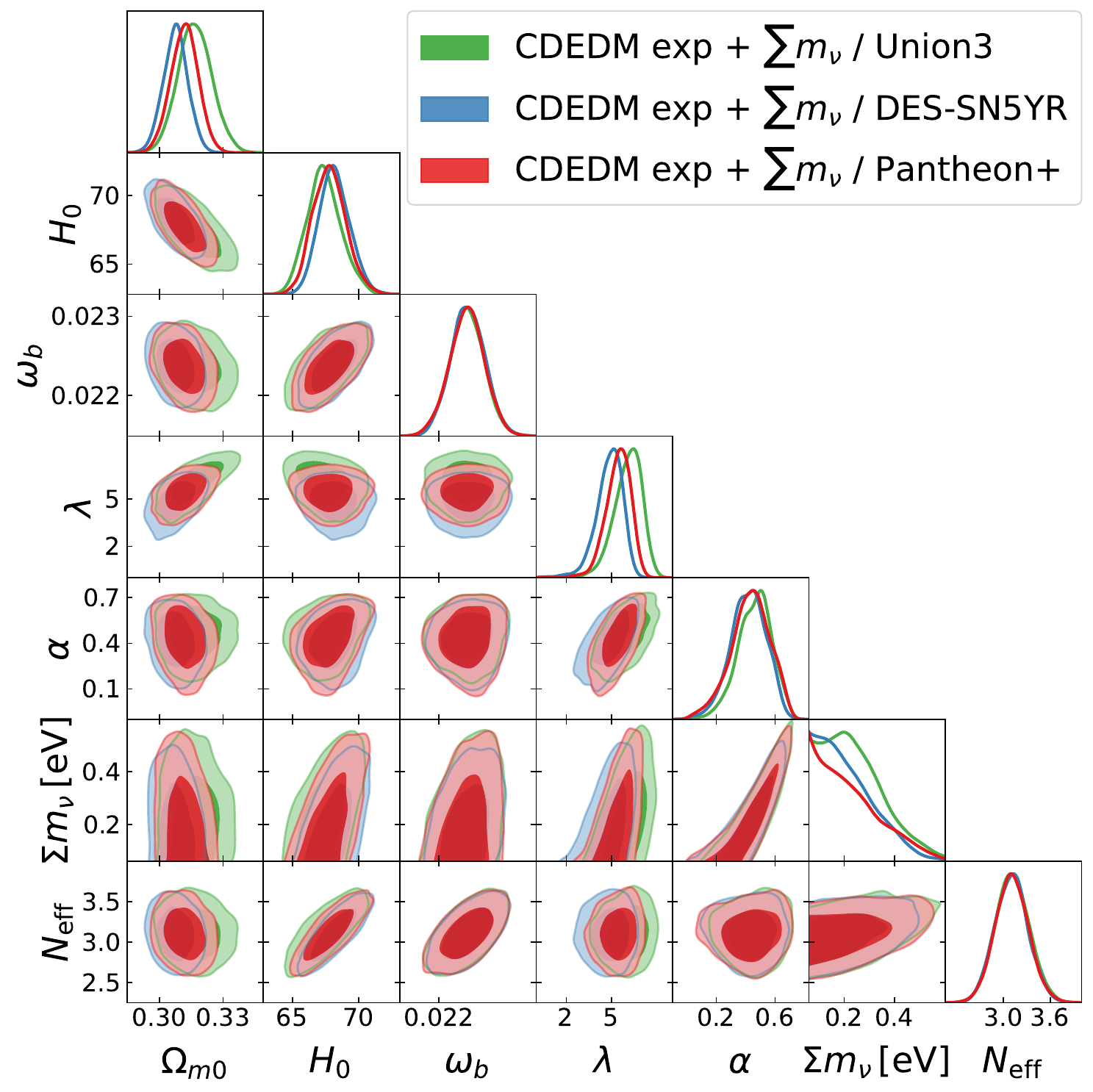}\hfill
\includegraphics[width=0.49\textwidth]{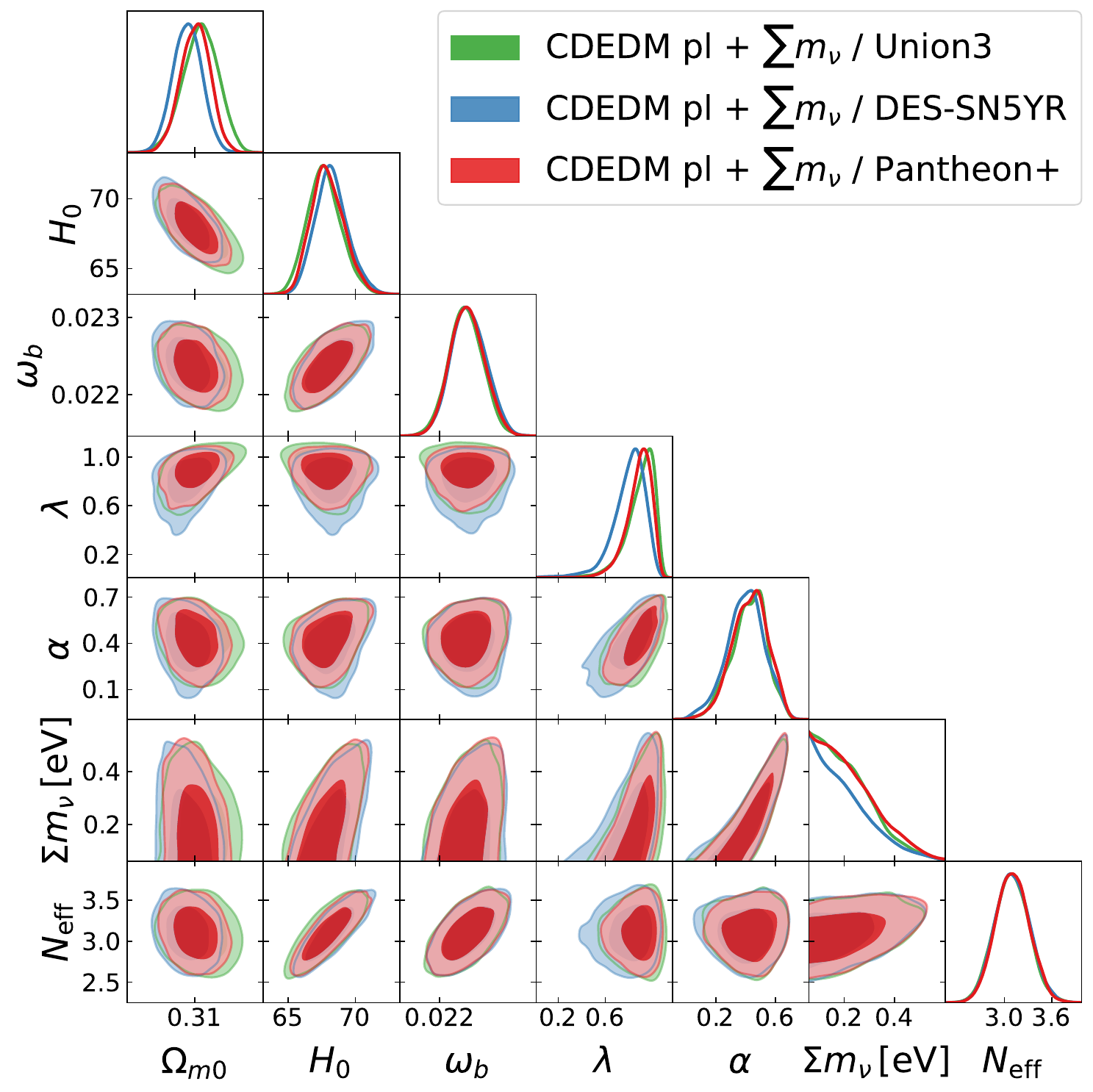}
\caption{Posterior distributions for the seven sampled parameters of the coupled model, for the exponential potential $V\propto e^{-\lambda\phi/\sqrt{8\pi}M_{\rm pl}}$ (left) and the inverse power law $V\propto\phi^{-\lambda}$ (right). Contours are $68\%$ and $95\%$, overlaid for the three supernova compilations: Union3 (green), DES-SN5YR (blue) and Pantheon$+$ (red).}
\label{Fig1}
\end{figure}

Table~\ref{TableI} and Fig.~\ref{Fig1} show that the inferred cosmological parameters remain stable across the three supernova compilations, whereas the potential parameters display the expected model dependence. For the exponential potential, we find a preference for $\lambda\simeq 5$-$6$, while for the inverse power-law potential, the preferred range shifts to $\lambda\simeq 0.8$-$0.9$, consistent with the different scaling of $\lambda$ in the two parametrizations. In both cases, we obtain $\alpha\simeq 0.4$-$0.5$. The agreement of $\Omega_{m0}$, $H_0$, $\omega_b$, $N_{\rm eff}$, and $\sum m_\nu$ across the supernova samples indicates that our coupling constraints are not driven by any single supernova compilation. It should be noted that throughout this work, in all tables, ``exp'' denotes the exponential potential and ``pl'' denotes the inverse power-law potential.

\begin{table}[t]
\caption{Marginalised constraints on the seven sampled parameters of the coupled model, for the three supernova compilations along with CMB distance-prior, RSD data, DESI DR2 BAO data, ACT lensing constraint and for both potentials.}
\label{TableI}
\centering
\setlength{\tabcolsep}{2pt}
\scriptsize
\begin{ruledtabular}
\resizebox{\linewidth}{!}{%
\begin{tabular}{llccccccc}
Data & $V(\phi)$ & $\Omega_{m0}$ & $H_0$ & $\omega_b$ & $N_{\rm eff}$
 & $\sum m_\nu$~[eV] & $\lambda$ & $\alpha$ \\
 & & & [km\,s$^{-1}$Mpc$^{-1}$] & & & ($95\%$ C.I.) & & \\
\hline
DES-SN5YR & exp & $0.3077^{+0.0055}_{-0.0060}$ & $68.13^{+1.22}_{-1.12}$ & $0.02238^{+0.00023}_{-0.00022}$ & $3.103^{+0.209}_{-0.215}$ & $[0.066,\,0.459]$ & $4.93^{+0.75}_{-0.92}$ & $0.423^{+0.121}_{-0.124}$ \\
Pantheon$+$ & exp & $0.3119^{+0.0063}_{-0.0067}$ & $67.77^{+1.25}_{-1.26}$ & $0.02237^{+0.00022}_{-0.00023}$ & $3.104^{+0.217}_{-0.210}$ & $[0.065,\,0.503]$ & $5.49^{+0.74}_{-0.87}$ & $0.438^{+0.133}_{-0.139}$ \\
Union3 & exp & $0.3166^{+0.0082}_{-0.0077}$ & $67.34^{+1.34}_{-1.26}$ & $0.02237^{+0.00022}_{-0.00022}$ & $3.112^{+0.223}_{-0.216}$ & $[0.069,\,0.508]$ & $6.16^{+0.80}_{-1.01}$ & $0.472^{+0.106}_{-0.125}$ \\
\hline
DES-SN5YR & pl & $0.3070^{+0.0059}_{-0.0060}$ & $68.17^{+1.28}_{-1.17}$ & $0.02238^{+0.00025}_{-0.00023}$ & $3.097^{+0.222}_{-0.214}$ & $[0.065,\,0.474]$ & $0.825^{+0.109}_{-0.148}$ & $0.405^{+0.122}_{-0.129}$ \\
Pantheon$+$ & pl & $0.3103^{+0.0058}_{-0.0062}$ & $67.88^{+1.31}_{-1.15}$ & $0.02237^{+0.00024}_{-0.00022}$ & $3.103^{+0.215}_{-0.214}$ & $[0.066,\,0.480]$ & $0.903^{+0.088}_{-0.122}$ & $0.436^{+0.112}_{-0.130}$ \\
Union3 & pl & $0.3125^{+0.0070}_{-0.0075}$ & $67.65^{+1.31}_{-1.21}$ & $0.02235^{+0.00023}_{-0.00022}$ & $3.092^{+0.217}_{-0.221}$ & $[0.066,\,0.464]$ & $0.928^{+0.089}_{-0.143}$ & $0.429^{+0.103}_{-0.133}$ \\
\end{tabular}%
}
\end{ruledtabular}
\end{table}

\begin{table}[t]
\caption{\label{TableII}Model comparison}
\begin{ruledtabular}
\begin{tabular}{llccccc}
Model & Data & $k$ & $\chi^2_\nu$ & $\Del\ln Z_{w_0w_a}$ & $\Del\ln Z_{\Lam}$ & $\Del$DIC \\
\hline
$w_0w_a$ & DES  & 7 & 0.8962 & $0$    & $-0.05$ &  $-6.3$ \\
exp      & DES  & 7 & 0.8963 & $2.96$ & $2.90$  &  $-6.9$ \\
pl       & DES  & 7 & 0.8964 & $2.50$ & $2.44$  &  $-6.3$ \\
\hline
$w_0w_a$ & Pan. & 7 & 0.9023 & $0$    & $2.28$  &  $-10.5$ \\
exp      & Pan. & 7 & 0.9023 & $2.64$ & $4.92$  &  $-10.8$ \\
pl       & Pan. & 7 & 0.9022 & $2.00$ & $4.28$  &  $-11.3$ \\
\hline
$w_0w_a$ & Un3  & 7 & 0.9113 & $0$    & $2.22$  &  $-10.3$ \\
exp      & Un3  & 7 & 0.9304 & $1.94$ & $4.16$  &  $-9.8$ \\
pl       & Un3  & 7 & 0.9374 & $0.73$ & $2.95$  &  $-8.0$ \\
\end{tabular}
\end{ruledtabular}
\end{table}

Table~\ref{TableII} summarizes the Bayesian model comparison. We quote $\Del\ln Z$ relative to $w_0w_a$CDM; since both $w_0w_a$CDM and CDEDM have $k=7$ parameters, are fit to the same data vector, and are sampled in the same way, the preference cannot be ascribed to extra model freedom. For CDEDM, we find $\Del\ln Z_{w_0w_a}=1.92$--$2.84$ for the exponential potential and $0.86$--$2.51$ for the inverse power-law potential, corresponding to Bayes factors $\exp(\Del\ln Z)\simeq 2$--$17$, with the strongest support from DES-SN5YR and the weakest from Union3. Relative to $\Lam$CDM, where the comparison does involve a genuine Occam penalty, the evidence differences remain positive, spanning $\Del\ln Z_{\Lam}=2.28$--$4.92$ across the three supernova compilations and the two potentials.

The DES-SN5YR results merit separate emphasis. Relative to $\Lam$CDM, the seven-parameter $w_0w_a$CDM model yields $\Del\ln Z=-0.05$, i.e. its modest improvement in fit is essentially cancelled by the Occam penalty for the two additional parameters. By contrast, the coupled model with the same parameter count is strongly preferred on the same data set, with $\Del\ln Z_{\Lam}=2.96$ for the exponential potential (and $2.50$ for the inverse power-law case). This indicates that, for DES-SN5YR, the coupling produces an expansion history that cannot be replicated by a purely kinematic $(w_0,w_a)$ parametrization.

The information criteria point in the same direction. Relative to $\Lam$CDM, we find $\Del\mathrm{DIC}=-6.3$ to $-11.3$ for CDEDM across the three supernova compilations (depending on the potential), which is comparable to the improvement obtained in $w_0w_a$CDM. At the best fit, the BAO contribution ($13$ points) is likewise similar in the two extended models: $\chi^2_{\rm BAO}=8.41$--$9.10$ for CDEDM and $7.79$--$9.24$ for $w_0w_a$CDM, compared to $11.21$--$12.29$ for $\Lam$CDM. Thus, BAO alone does not meaningfully distinguish CDEDM from a kinematic $(w_0,w_a)$ extension; both mainly improve on $\Lam$CDM at the background level, so any remaining discriminating power may come from growth-related observables.

Overall, Table~\ref{TableII} shows that CDEDM is relatively preferred by the Bayesian evidence: it may prefer over $w_0w_a$CDM despite having the same parameter count, and it also improves over $\Lam$CDM even after the Occam penalty is accounted for. The preference is strongest for DES-SN5YR and weakest for Union3, while the DIC trends are broadly consistent with the evidence-based ranking.

\subsection{Comparison of Equation-of-state}
\begin{figure}
\centering
\includegraphics[width=0.98\textwidth]{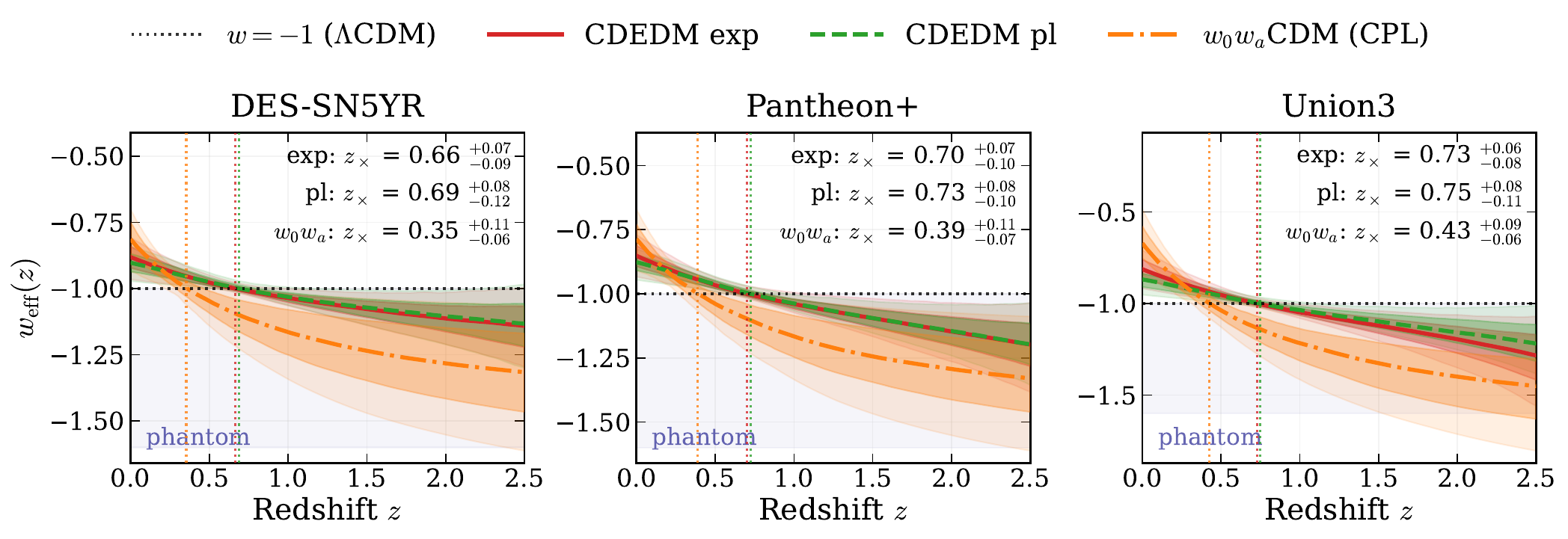}
\caption{Effective equation of state $w_{\rm eff}(z)$ for both CDEDM potentials and $w_0w_a$CDM, for the three supernova compilations. Bands are $68\%$ and $95\%$ credible intervals}
\label{Fig2}
\end{figure}

Figure~\ref{Fig2} plots the effective equation of state $w_{\rm eff}(z)$ from Eq.~\eqref{5}. For both scalar potentials, all posterior samples cross the phantom divide, reaching a median minimum $w_{\rm eff}=-1.127$, while the underlying field remains non-phantom with $w_\phi\ge-1$ at all redshifts, as required by Eq.~\eqref{5}. The crossing redshift is $z_\times=0.66^{+0.07}_{-0.09}$, $0.70^{+0.07}_{-0.10}$, and $0.73^{+0.06}_{-0.08}$ for the exponential potential, and $0.69^{+0.08}_{-0.12}$, $0.73^{+0.08}_{-0.10}$, and $0.75^{+0.08}_{-0.11}$ for the inverse power-law potential, for DES-SN5YR, Pantheon$+$, and Union3, respectively. This occurs systematically later than in $w_0w_a$CDM, which crosses at $z_\times=0.355^{+0.105}_{-0.065}$, $0.391^{+0.115}_{-0.071}$, and $0.426^{+0.086}_{-0.060}$, and with a noticeably steeper transition driven by the kinematic parameter $w_a$ rather than an interaction.

The behavior in Fig.~\ref{Fig2} is driven by the dark-sector coupling $\alpha$, which is consistently bounded away from zero for every data set and for both scalar potentials. The marginalized medians are $\alpha=0.423^{+0.121}_{-0.124}$, $0.438^{+0.133}_{-0.139}$, and $0.472^{+0.106}_{-0.125}$ for the exponential potential, and $0.405^{+0.122}_{-0.129}$, $0.436^{+0.112}_{-0.130}$, and $0.429^{+0.103}_{-0.133}$ for the inverse power-law potential (DES-SN5YR, Pantheon$+$, and Union3, respectively). The corresponding $95\%$ lower bounds are $\alpha>0.206$, $0.187$, $0.254$ (exponential) and $\alpha>0.177$, $0.224$, $0.209$ (inverse power law).

We report lower bounds rather than a ``$\sigma$ from zero'' because the baseline prior is one-sided, $\alpha\in[0,1]$: obtaining $w_{\rm eff}<-1$ requires $\alpha>0$, so the null value lies on the prior boundary and a Gaussian-equivalent significance is not well-defined. Using a Savage--Dickey density ratio with the same prior, $\alpha\sim U[0,1]$, we obtain $\mathrm{BF}_{10}=9.3$, $8.1$, and $31.8$ (exponential) and $7.9$, $31.2$, and $9.2$ (inverse power law), i.e.\ $\ln\mathrm{BF}_{10}=2.1$--$3.5$ and $2.1$--$3.4$. Since Savage--Dickey Bayes factors scale inversely with the assumed prior width, these values should be interpreted only in conjunction with the stated prior range. On the Jeffreys scale~\cite{Jeffrey1961}, they correspond to strong to very strong evidence for a nonzero coupling, with the strongest support in the Union3+exponential case. A robustness check with a two-sided prior on $\alpha$ is presented in Appendix~\ref{sec:twosided}.

In summary, Fig.~\ref{Fig2} shows that the effective equation of state $w_{\rm eff}(z)$ crosses below $-1$ for all three supernova compilations, while the underlying scalar remains canonical ($w_\phi\ge -1$). The preferred positive coupling may provide a consistent dynamical mechanism for this phantom crossing and yields a later, smoother transition than the steeper crossing found in the purely kinematic $w_0w_a$CDM parametrization. 

\subsection{The running dark-matter mass at recombination}\label{sec:recomb}

\begin{figure}[t]
\centering
\includegraphics[width=0.98\textwidth]{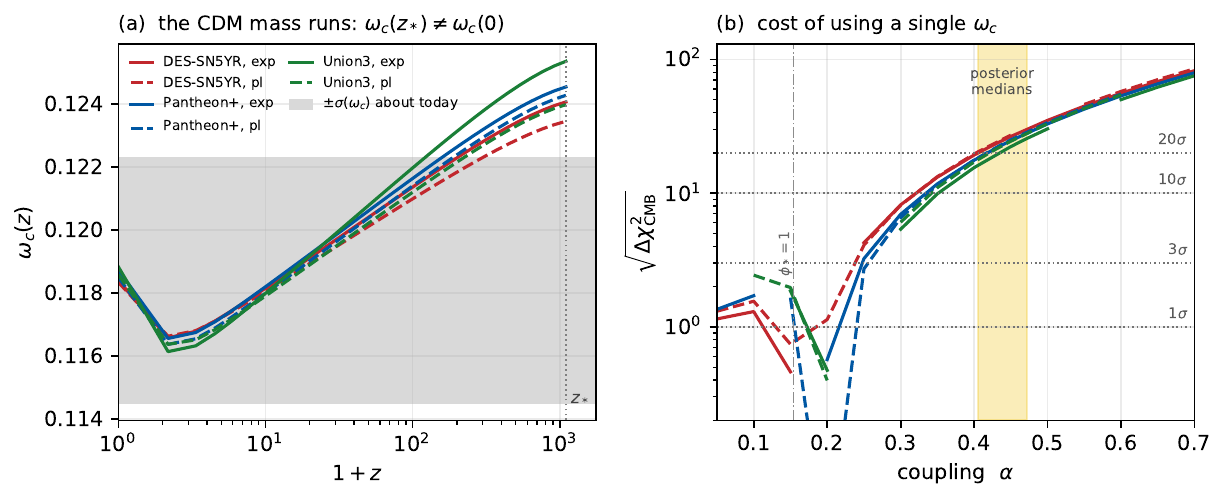}
\caption{\label{Fig3}The running dark-matter mass and the compressed CMB. \textit{(a)} $\om_c(z)$ from the present epoch to last scattering at each chain's posterior median, against the $\pm\sigma(\om_c)$ band of the compressed prior about the present-day value. \textit{(b)} The compressed-CMB penalty incurred by integrating $c_s/H$ with $\om_c$ frozen at its present value instead of with the running mass, at otherwise fixed cosmology. The dash-dotted line marks $\phi_*=1$, where the field value at recombination equals its present value and the sign of the error reverses.}
\end{figure}

The coupling makes the dark-matter mass evolve with the scalar field, so the comoving cold-dark-matter density is not constant in time. In the gauge $\phi_0\equiv1$, the mass factor is
\be
\frac{f(\phi)}{f_0}=\exp\![\beta(\phi-1)].
\ee
Consequently, the physical CDM density $\om_c$ at recombination differs from its present-day value. For the models considered here, the field has $\phi_*=1.523$--$1.571$ at $z_*$ and evolves to $\phi=1$ today, implying a larger dark-matter mass at recombination by $4.4\%$--$5.5\%$. This corresponds to $\om_c(z_*)=0.1235\text{--}0.1257$ and $\om_c(0)=0.1183\text{--}0.1191$, as reported in Table~\ref{TableIII}. Figure~\ref{Fig3}(a) shows the full evolution, which is non-monotonic: $\om_c(z)$ dips slightly below its present value around $z\simeq2$ before rising toward recombination. There is therefore no unique epoch at which one can simply quote ``the'' $\om_c$ of the model.

The apparent significance of the recombination--today shift depends on which uncertainty is used for comparison. Relative to the \textit{Planck} uncertainty $\sigma(\om_c)=9.1\times10^{-4}$ (with $\sigma$ denoting the standard deviation)~\cite{ATLAS}, the shift corresponds to about $5.8$--$7.3\sigma$. Using instead the conditional uncertainty of the compressed prior, $1.0\times10^{-3}$, gives roughly $5\sigma$, whereas the marginalized uncertainty of the same prior, $3.9\times10^{-3}$, gives only about $1.3\sigma$. We do not interpret these as direct constraints on $\om_c$, because the compressed likelihood does not constrain $\om_c$ independently; its main information is on the acoustic scale. If one compares the acoustic scale computed with $\om_c$ frozen at its present value to that computed with the running mass, the corresponding penalty is
\be
\sqrt{\Del\chi^2}\;\equiv\;\frac{|\Del\ell_A|}{\sigma_{\rm cond}(\ell_A)},\qquad \sigma_{\rm cond}(\ell_A)=0.117,
\ee
which gives $\sqrt{\Del\chi^2}=20.8$ for the DES-SN5YR posterior median with the exponential potential, and $20.8$--$25.8$ across all six chains.

It is worth emphasizing, what this number means. The coupled model itself is not in tension with the compressed acoustic-scale constraint: it predicts $\ell_A=301.74$, only $0.06\sigma$ from the compressed mean $\ell_A=301.73$. The large $\sqrt{\Del\chi^2}$ above therefore reflects the difference between two treatments of the dark-matter evolution, rather than a poor fit of the coupled model to $\ell_A$.

The dependence on the coupling is also sharp and, crucially, non-monotonic (Fig.~\ref{Fig3}(b)). For weak coupling, $\alpha\lesssim0.15$--$0.20$, the shooting solution gives $\phi_*<1$, so the potential largely controls the field evolution. The dark matter is then lighter at recombination, and holding $\om_c$ fixed at its present-day value biases the acoustic-scale calculation in the opposite direction. At the crossing point, $\phi_*=1$ (the dash-dotted line), the error crosses through zero, so a constant $\om_c$ approximation happens to work there. An analysis calibrated only near this crossing could therefore find almost no effect and mistakenly conclude that the running mass is negligible. Away from the crossing the mismatch grows rapidly, reaching $\sqrt{\Del\chi^2}\simeq3$ by $\alpha\simeq0.28$ and $\simeq10$ by $\alpha\simeq0.36$, both still below the couplings preferred by our posterior samples.

\begin{table}[t]
\caption{\label{TableIII}Running of the dark-matter mass, and the scalar energy density, at recombination, evaluated at each chain's posterior median. Here $\phi_*$ is the field value at $z=1100$ with $\phi_0\equiv1$, and $f(\phi_*)/f_0$ is the dark-matter mass relative to its present value. The shift column is $[\om_c(z_*)-\om_c(0)]$ in units of the \textit{Planck} uncertainty $\sigma(\om_c)=9.1\times10^{-4}$, not of the compressed prior's own error. The last two columns give the fractional change in the sound horizon when $\om_c$ is instead held at its present value, and the resulting compressed-CMB penalty $\sqrt{\Del\chi^2}=|\Del\ell_A|/
\sigma_{\rm cond}(\ell_A)$.}
\begin{ruledtabular}
\setlength{\tabcolsep}{2pt}
\scriptsize
\begin{tabular}{llccccccc}
Data & $V$ & $\phi_*$ & $f(\phi_*)/f_0$ & $\om_c(z_*)$ &
$\Del\om_c/\sigma_{\rm Planck}$ & $\Om_\phi(z_*)$ & $\Del r_s/r_s$ &
$\sqrt{\Del\chi^2}$ \\
\hline
DES-SN5YR & exp & $1.554$ & $1.0479$ & $0.1241$ & $+6.2$ & $4.9\times10^{-4}$ & $+0.88\%$ & $22.5$ \\
DES-SN5YR & pl & $1.523$ & $1.0432$ & $0.1234$ & $+5.6$ & $4.5\times10^{-4}$ & $+0.79\%$ & $20.3$ \\
Pantheon$+$ & exp & $1.547$ & $1.0489$ & $0.1245$ & $+6.4$ & $5.3\times10^{-4}$ & $+0.89\%$ & $22.9$ \\
Pantheon$+$ & pl & $1.547$ & $1.0488$ & $0.1243$ & $+6.4$ & $5.2\times10^{-4}$ & $+0.89\%$ & $22.9$ \\
Union3 & exp & $1.568$ & $1.0549$ & $0.1254$ & $+7.2$ & $6.1\times10^{-4}$ & $+0.99\%$ & $25.6$ \\
Union3 & pl & $1.519$ & $1.0455$ & $0.1240$ & $+5.9$ & $5.1\times10^{-4}$ & $+0.83\%$ & $21.3$ \\
\end{tabular}
\end{ruledtabular}
\end{table}

There is, however, an important qualification to this comparison. In a full analysis, we would not keep all other cosmological parameters fixed while changing how the dark-matter evolution is treated. One would refit the model, allowing part of the shift in $\ell_A$ to be absorbed by other parameters, which would reduce the residual $\chi^2$ difference. That does not make the effect irrelevant: in a coupled model the coupling $\alpha$ can naturally compensate changes in the acoustic scale, so neglecting the running mass can directly bias the inferred coupling.

For this reason, we account for the running mass consistently within the different pieces of the compressed likelihood. The shift parameter $R$ and the acoustic scale $\ell_A$ are evaluated using the matter density at recombination, $\Omega_m(\phi_*)$. The sound horizons $r_s$ and $r_d$ are obtained by directly integrating $c_s/H$ out to $z=10^6$ while updating the dark-matter mass along the integration; no single constant choice of $\om_c$ reproduces the same integral. For the linear transfer function, we use the equality-era value of $\om_c$, since the turnover scale is controlled by $k_{\rm eq}\sim a_{\rm eq}H(a_{\rm eq})$, as in standard analytic transfer-function treatments~\cite{BBKS1986,HuSugiyama1996,EisensteinHu1998}.

This last step should be regarded as an approximation. The scalar energy density behaves rather differently. At recombination we find $\Omega_\phi(z_*)=4.9\times10^{-4}\text{--}6.4\times10^{-4}$, well below current early-dark-energy limits. The scalar is also almost entirely kinetic at that epoch, with $w_\phi\simeq+1$. The coupling source term, $3H_0^2\Omega_{c0}(1+z)^3\frac{f_{,\phi}}{f_0}$, drives $\dot\phi$ toward the quasi-static attractor, while the potential energy remains extremely small, $\Omega_V\simeq5.3\times10^{-10}\text{--}8.3\times10^{-10}$ across the model--data combinations. At higher redshift this contribution decreases approximately as $(1+z)^{-2}$, because $d\phi/dN$ falls as $1/(1+z)$ once radiation domination sets in. It therefore peaks around recombination and is never important in the sound-horizon integral: its net effect shifts $r_s$ by $\Delta r_s/r_s=-(1.2\times10^{-4}\text{--}1.6\times10^{-4})$, and changes the acoustic scale by $\Delta\ell_A=+0.035\text{--}+0.047$, i.e. about $0.39$--$0.53\sigma(\ell_A)$. Thus, although the potential energy is tiny, its impact on $\ell_A$ is still large enough to be worth reporting explicitly.

Finally, our treatment of the linear transfer function is limited. The emulator was trained on $\Lambda$CDM models with constant $\om_c$, so providing it with the equality-era value of $\om_c$ is only a leading-order correction. In the coupled model, the transfer function depends on the full evolution of the dark-matter density through equality, not on a single value evaluated at one epoch. A Boltzmann calculation that explicitly includes the coupled-quintessence sector would provide the appropriate treatment and would allow the residual error from this approximation to be quantified. We therefore regard the present prescription as a controlled approximation, not as a substitute for a full coupled-Boltzmann analysis.

In summary, the coupling induces a time-dependent dark-matter mass, so quantities that enter the compressed CMB likelihood (notably $r_s$, $r_d$, $R$, and $\ell_A$) must be evaluated using the running dark-matter density rather than a single fixed $\om_c$; otherwise one can incur a large, coupling-dependent bias even when the coupled model itself remains consistent with the acoustic-scale constraint.

\subsection{Residual check}

\begin{figure}
    \centering
    \includegraphics[width=0.95\textwidth]{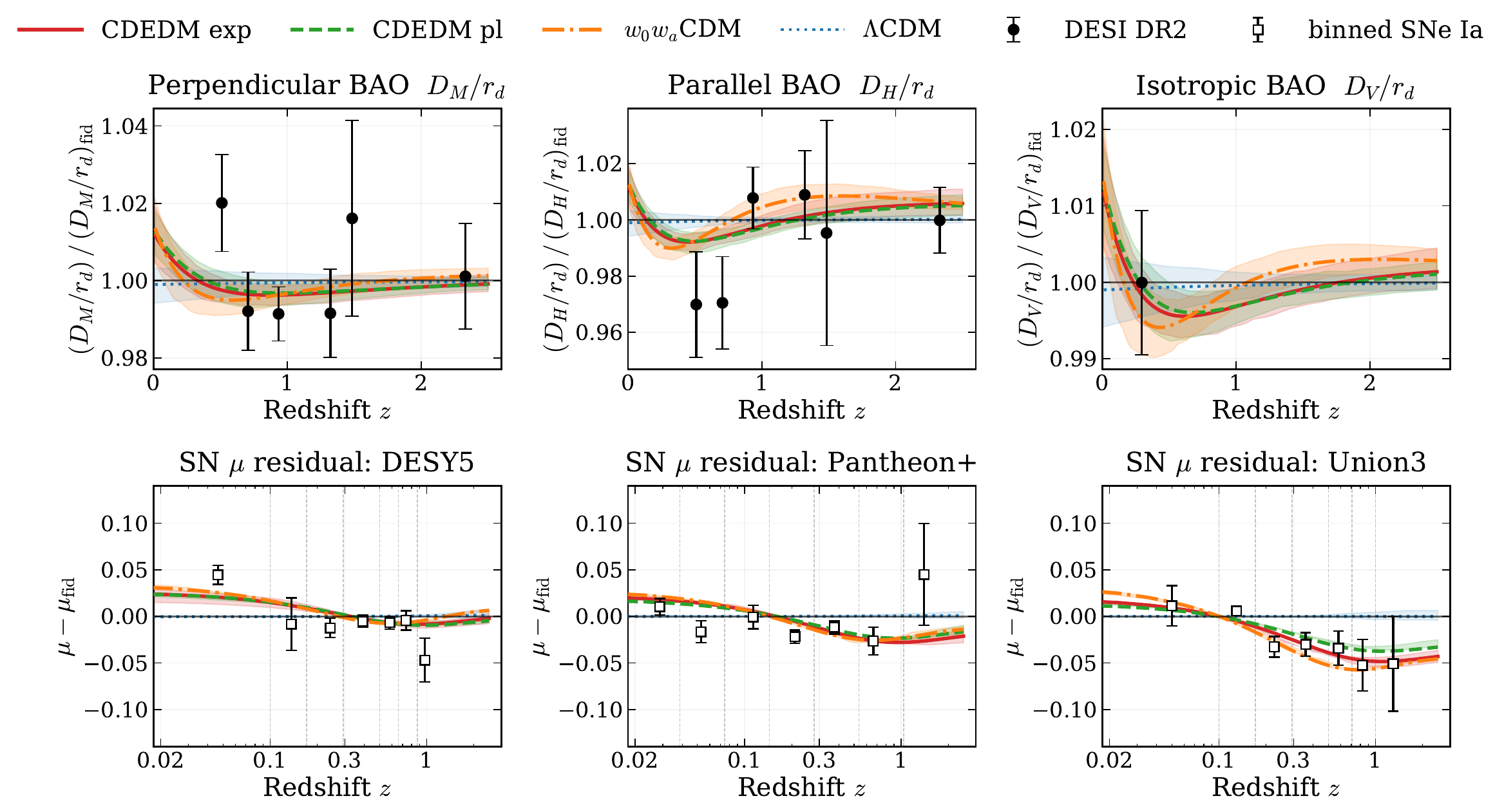}
    \caption{Residuals relative to dataset-specific $\Lambda$CDM fiducial models. \textit{Top:} BAO distance-measure residuals ($D_M/r_d$, $D_H/r_d$, $D_V/r_d$) for DESI~DR2~\cite{DESI2025}. \textit{Bottom:} SN~Ia distance-modulus residuals for DES-SN5YR, Pantheon$+$, and Union3. Shaded bands indicate $68\%$ credible intervals.}
    \label{Fig4}
\end{figure}

Figure~\ref{Fig4} shows residuals with respect to a dataset-specific $\Lambda$CDM fiducial for both BAO distance measures and SN~Ia distance moduli. In the BAO panels (top row), CDEDM with an exponential potential, CDEDM with an inverse power-law potential, and $w_0w_a$CDM all track the DESI~DR2 points across $D_M/r_d$, $D_H/r_d$, and $D_V/r_d$. They reproduce the characteristic pattern around $z\simeq 0.5$: an upward fluctuation of order $\sim\!2\%$ in $D_M/r_d$ accompanied by a compensating downward shift in $D_H/r_d$, suggestive of a low-to-intermediate redshift departure from the $\Lambda$CDM distance--redshift relation. In the SN~Ia panels (bottom row), the same models capture the coherent drift to negative distance-modulus residuals for $z\gtrsim 0.3$ seen in DES-SN5YR, Pantheon$+$, and Union3. The inverse power-law potential follows the high-$z$ slope most closely, while the exponential potential lies slightly lower over $0.3\lesssim z\lesssim 1$ but remains consistent with the binned measurements within the $68\%$ posterior bands. By construction, $\Lambda$CDM yields nearly flat residuals and therefore cannot account for the shared negative trend in the supernova compilations. Overall, Fig.~\ref{Fig4} provides a compact visual check that the dynamical dark-energy fits simultaneously respect the DESI BAO distances and the supernova residual pattern that drives the preference away from $\Lambda$CDM.

\subsection{Dark-sector energy exchange}

\begin{figure}
    \centering
    \includegraphics[width=0.95\textwidth]{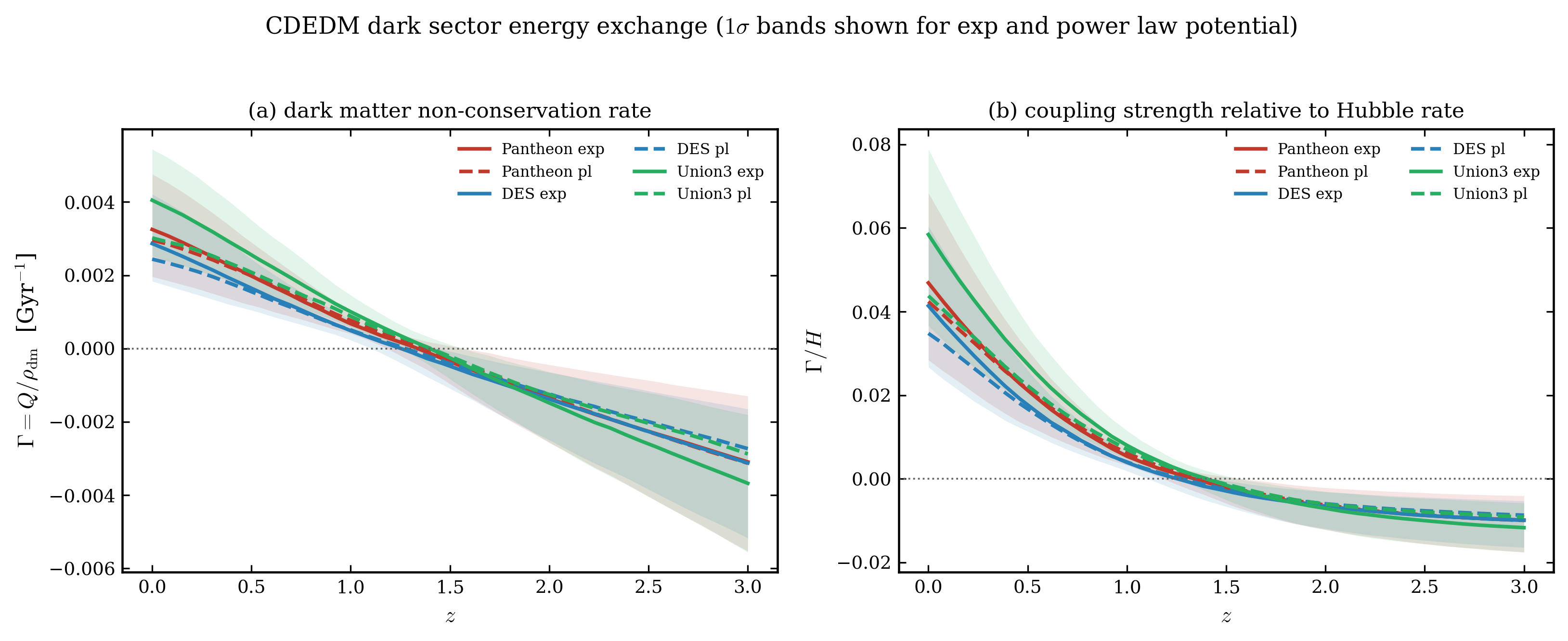}
    \caption{Late-time evolution of the dark-sector energy exchange. The left panel shows the dark-matter nonconservation rate $\Gamma=Q/\rho_{\rm DM}$, while the right panel shows the interaction strength relative to the Hubble rate, $\Gamma/H$. Results are shown for Pantheon$+$ (red), DES-SN5YR (blue), and Union3 (green), with shaded regions denoting the corresponding $68\%$ credible intervals.}
    \label{Fig5}
\end{figure}

The inferred dark-sector interaction shows a coherent redshift dependence across the three supernova compilations and for both scalar-field potentials. In Fig.~\ref{Fig5}, the dimensionless transfer rate $\Gamma/H$ is positive at low redshift (of order $\mathcal{O}(10^{-2})$ at $z=0$) and decreases toward higher redshift, crossing $\Gamma/H=0$ at $z\simeq 1.3$ and becoming mildly negative at earlier times. The corresponding dark-matter nonconservation rate, $\Gamma=Q/\rho_{\rm DM}$, follows the same qualitative evolution. Overall, the interaction is therefore concentrated at late times, with an inferred magnitude that is small at earlier epochs.

\subsection{The coupling—neutrino-mass degeneracy}
\label{sec:degeneracy}

A key signature of the coupled dark-sector scenario is a pronounced degeneracy between the coupling strength $\alpha$ and the summed neutrino mass $\sum m_\nu$. Both parameters affect the growth of structure but in opposite directions: the scalar-mediated fifth force associated with nonzero $\alpha$ enhances clustering, while neutrino free streaming suppresses it. As a result, growth-sensitive data admit a characteristic compensation between $\alpha$ and $\sum m_\nu$, which we quantify below and compare with an analytic estimate.

\begin{figure}[t]
\includegraphics[width=0.98\textwidth]{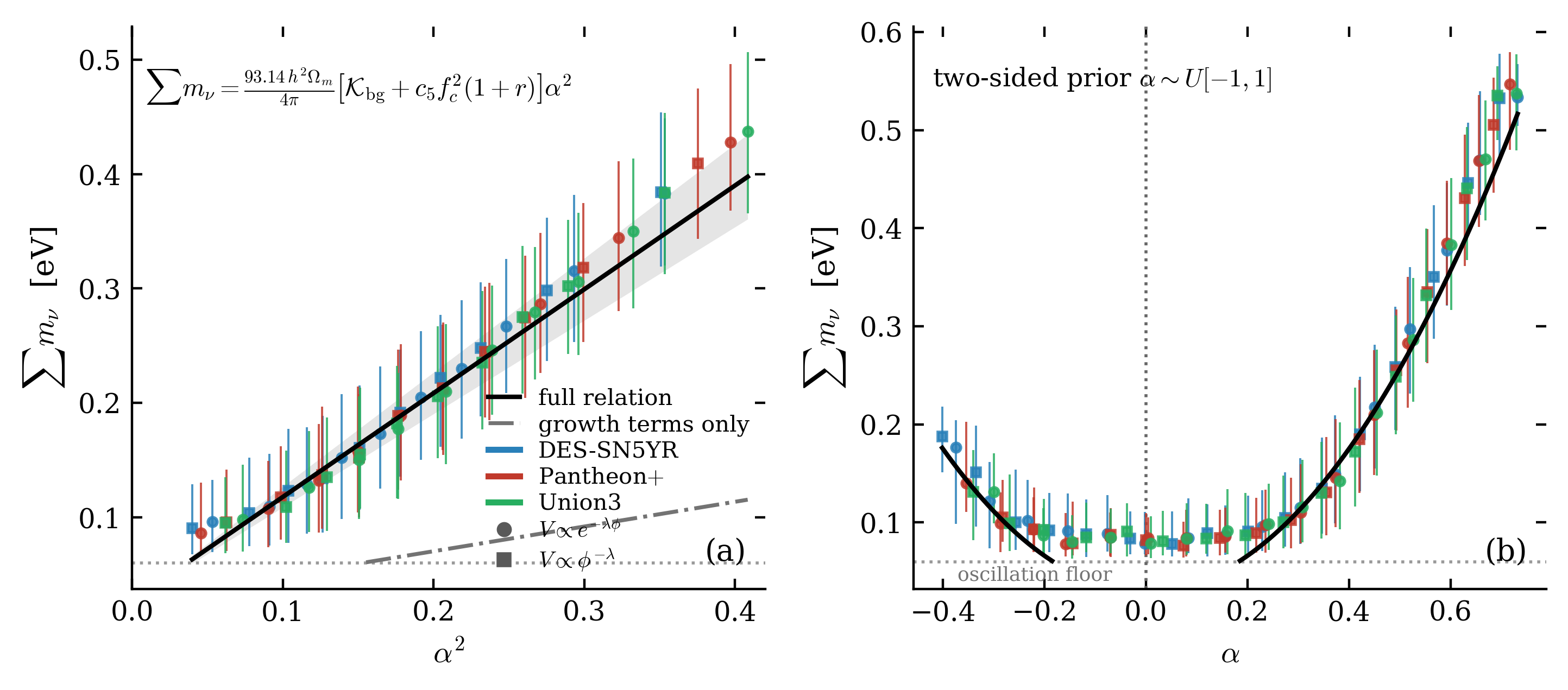}
\caption{\label{Fig6}Coupling--neutrino-mass degeneracy for the three supernova compilations and both scalar potentials. Panel (a) compares the posterior relation with the analytic prediction of Eq.~(\ref{12}), the dash-dotted curves retain only the fifth-force and drag contributions. Panel (b) tests the quadratic $\alpha$ dependence using a two-sided prior.}
\end{figure}

Table~\ref{TableI} shows that the coupled model prefers a relatively large neutrino mass, with posterior medians $\sum m_\nu\simeq 0.18$--$0.23\,\mathrm{eV}$ across the three supernova compilations (for both scalar-field potentials). The coupling strength $\alpha$ is in turn tightly correlated with $\sum m_\nu$ (Pearson $r\simeq 0.81$--$0.85$; see Fig.~\ref{Fig6}). Physically, the two parameters compensate in the growth sector: the scalar-mediated fifth force associated with nonzero coupling enhances structure growth, while neutrino-free streaming suppresses it.

To obtain a transparent estimate of the growth-sector degeneracy, we work deep in matter domination and assume a constant CDM--scalar coupling $\alpha$. We neglect both the scalar potential and radiation and treat the scalar-mediated fifth force as effectively long-ranged so that the enhancement of the CDM gravitational interaction is scale-independent. This approximation is well satisfied here: with a Compton wavelength $\bar\lambda_C=3234\,\mathrm{Mpc}$ (where $\bar\lambda_C=\hbar/(m_\phi c)$ and $m_{\phi}=\sqrt{V_{\rm eff,\phi\phi}}$), the corresponding transition scale is $k\simeq4.5\times10^{-4}\,h\,\mathrm{Mpc}^{-1}$, far below the observable range, and hence $G_{\rm eff}/G_N=1+2\beta^2$ may be taken as scale independent. For the posterior-median CDEDM solution, the growth ratio $\mathcal{R}(k,z)\equiv D_{\rm CDEDM}/D_{w_0w_a}$ varies by only $0.001\%$ across $0.005\leq k\leq0.30\,h\,\mathrm{Mpc}^{-1}$, indicating no appreciable scale-dependent signature. Baryons remain uncoupled, and neutrinos are treated as a freely streaming (non-clustering) component on the scales relevant to the growth constraint.\\

Only cold dark matter feels the fifth force, so the perturbations cannot be treated as a single effective fluid. Baryons contribute to (and respond to) the gravitational potential, but they are not directly coupled to the scalar, so we must evolve the two species together,
\ben
\begin{aligned}
\del_c'' + \tfrac12\del_c'
  &= \tfrac32\big[f_c(1+2\beta^2)\del_c + f_b\del_b\big],\\
\del_b'' + \tfrac12\del_b'
  &= \tfrac32\big[f_c\del_c + f_b\del_b\big],
\end{aligned}
\qquad \beta\equiv\frac{\al}{\sqrt{8\pi}},
\label{11}
\een
where $f_c\equiv\Om_c/\Om_m$ and $f_b\equiv 1-f_c$. Writing $\del_c\propto A\,a^{p}$ and $\del_b\propto B\,a^{p}$ and expanding about the uncoupled growing mode ($p=1$ with $A=B$), the difference of the two equations fixes the CDM-baryon offset, $A-B=2f_c\beta^2$. Substituting back then yields the growing-mode index,
\ben
p_g = 1 + \tfrac65 f_c^{2}\beta^{2} + \mathcal{O}(\beta^4).
\label{11b}
\een
The coupling therefore enters twice, through the modified CDM source term and through the CDM-baryon offset, so the leading correction scales as $f_c^2$. We verify this numerically by sweeping $f_c$ from $0.72$ to $0.92$ at fixed $\Om_m$ and isolating the fifth-force response in the integrator: the measured scaling is $f_c^{1.970}$, with a spread of $0.004$ across all eighteen $f\sigma_8$ bins. The remaining $\simeq 1.5\%$ shortfall from an exact square may reflect the mild redshift drift of $f_c(z)$ induced by $\rho_c\propto f(\phi)a^{-3}$.

Massive neutrinos modify the same growth index by reducing the clustering source below their free-streaming scale. In that regime, they contribute to the background density but not to the Poisson source, so the growing mode of the total matter perturbation approximately satisfies
$\del''+\tfrac12\del'-\tfrac32(1-f_\nu)\del=0$,
with $f_\nu\equiv\sum m_\nu/(93.14\,h^2\Om_m)$, and hence $p_\nu=1-\tfrac35 f_\nu$. Matching this suppression to the fifth-force enhancement in Eq.~(\ref{11b}) gives the neutrino mass that would mimic the fifth-force contribution alone,
\be
\sum m_\nu \big|_{\rm 5th} \;\simeq\; \frac{f_c^{2}}{4\pi}\,
\big(93.14\,h^2\Om_m\big)\,\al^{2}.
\label{12a}
\ee
As emphasized, Eq.~(\ref{12a}) isolates only the fifth-force channel and therefore does not capture the full measured degeneracy.

Beyond the fifth force, the coupling affects growth through two additional routes. First, because CDM is not separately conserved, $d\ln\rho_{\rm DM}/dN=-3+\beta\phi'$, the non-conservation rate $\epsilon\equiv\beta\phi'=d\ln f/dN$ enters the CDM perturbation equation as an effective drag term. Second, our gauge choice $\phi_0\equiv 1$ ties the initial field value $\phi_i$ to $\alpha$, so changing the coupling modifies $\rho_{\rm DM}\propto f(\phi)a^{-3}$ and therefore the background expansion history even before perturbations are considered. \\

We isolate the three channels or contributions (background expansion response, fifth force, and perturbation drag) by switching off the fifth force and the drag independently inside the integrator and remeasuring the growth-sector slope each time. Table~\ref{TableIV} summarizes the decomposition. With both perturbation-level mechanisms disabled, the background response accounts for $74.8\%$--$77.6\%$ of the slope; the fifth force contributes roughly one half of the full effect, while the drag offsets it by about one third. These fractions are stable across supernova compilations and for both potentials. Figure~\ref{Fig6}(a) makes the same point at the level of the posterior relations: the dash-dotted curve, which retains only the growth-equation terms, is much too shallow to reach the data, and the remaining gap to the full relation is supplied by the background channel.

The perturbation drag (set by the non-conservation rate $\epsilon\equiv\beta\phi'$) opposes the fifth force with a ratio in the range $-0.603$ to $-0.635$. The matter-era estimate $\epsilon/(3f_c\beta^2)=-0.66$ agrees at the $\simeq 6\%$ level, indicating that the cancellation is driven by the modified growth index rather than by a late-time amplitude shift. If we disable each contribution while holding the background fixed, then at $\alpha=0.414$ (DES-SN5YR, exponential potential) we find
\ben
\del\ln\sigma_8\big|_{\rm 5th} = +0.0354, \qquad
\del\ln\sigma_8\big|_{\epsilon} = -0.0183,
\label{13}
\een
so the drag cancels about half of the fifth-force enhancement in $\sigma_8$ (ratio $\approx-0.52$). For $f\sigma_8$, which is the quantity constrained by redshift-space distortions, the cancellation becomes redshift-dependent, increasing from $35\%$ at $z=0.2$ to $59\%$ at $z=1.8$, because the two effects enter different factors of the product.

All three channels scale predominantly as $\alpha^2$, but with different levels of accuracy. For the fifth force the quadratic scaling is exact (Eq.~(\ref{12a})); numerically, sweeping $\alpha=0.05$--$0.8$ we find that $\del\ln\sigma_8|_{\rm 5th}/\beta^2$ is constant to within $0.5\%$, consistent with the analytic expectation. The drag contribution is only approximately quadratic. With the gauge choice $\phi_0\equiv1$ one has
$\int \epsilon\,dN=\ln[f(\phi_0)/f(\phi_i)]=\beta(1-\phi_i)$ exactly, but $\phi_i$ depends affinely rather than proportionally on $\beta$ because the uncoupled field already rolls under its potential. As a result, a residual linear-in-$\beta$ piece remains; at the best fit it is about one fifth of the quadratic term. Over the parameter range supported by the posteriors, this linear correction is effectively absorbed into the fitted quadratic coefficient, which is why we characterize the drag through its measured ratio to the fifth-force response rather than attempting an independent analytic derivation.\\

The background trajectory does not admit a closed-form expression because the initial value $\phi_i$ is fixed implicitly: it must be tuned by a shooting procedure so that the chosen gauge condition $\phi_0\equiv 1$ is satisfied. For this reason, instead of attempting an analytic derivation, we extract the required relation numerically and use dimensional analysis to determine its functional form.

The quantity we want to predict is the total neutrino mass $\sum m_\nu$, which has dimensions of energy (eV). In the setup at hand, the only available dimensionful scale is the standard neutrino-density conversion factor
$\mathcal{D}\equiv 93.14\,h^2\Om_m$ (in eV), built from the dimensionless Hubble parameter $h$ and the matter fraction $\Om_m$. Consequently, any proportionality between $\sum m_\nu$ and the coupling parameter $\al$ must be linear in $\mathcal{D}$ and otherwise involve only dimensionless combinations. Since $\al$ is dimensionless, the leading dependence is quadratic, and we parametrize the result as
\be
\sum m_\nu \;=\; \frac{\mathcal{D}}{4\pi}\,\mathcal{K}\,\al^2,
\qquad
\mathcal{K}=\mathcal{K}_{\rm bg}+\mathcal{K}_{\rm 5th}+\mathcal{K}_{\rm drag},
\label{12b}
\ee
where the factor $1/(4\pi)$ is written explicitly by convention, and $\mathcal{K}$ is a pure number (dimensionless slope) that collects all remaining dependence on the dimensionless inputs. We further decompose $\mathcal{K}$ into additive contributions from the background evolution ($\mathcal{K}_{\rm bg}$), the fifth-force effect ($\mathcal{K}_{\rm 5th}$), and the drag term ($\mathcal{K}_{\rm drag}$). Thus $\mathcal{K}$ can depend only on $f_c$, $\Om_m$, $h$, and (through the background solution) the choice of potential.

Away from the best-fit point, $\mathcal{K}$ cannot be robustly inferred directly from the data: when we profile $\chi^2_{\rm growth}$ over $\sum m_\nu$, shifting the background parameters tends to push the minimum to the lower prior bound on $\sum m_\nu$. To avoid this prior-dominated behavior, we work purely in model space and define the degeneracy by requiring the predicted growth observable $f\sigma_8(z)$ to match the uncoupled model in each of the eighteen redshift-space bins. In this construction neither the growth data nor the neutrino-mass prior are used; this is appropriate because Eq.~(\ref{12a}) is intended as a relation within the model, not as the outcome of a particular fit. We evaluate this condition on a grid of twenty-eight points spanning $f_c=0.757$--$0.912$, $\Om_m=0.26$--$0.35$, and $h=0.63$--$0.72$ for both choices of potential, and we repeat the calculation with the fifth-force and drag terms switched off separately.

Holding the potential slope $\lambda$ fixed, we find that the fifth-force contribution is well described by
$\mathcal{K}_{\rm 5th}=c_5\,f_c^2$,
with $c_5$ varying by only $3.3\%$ over a $20\%$ scan in $f_c$. This provides an independent check of the predicted $f_c^2$ scaling, extracted here from the slope coefficient rather than from the growth response itself. The drag contribution follows the same scaling with an approximately constant ratio,
$\mathcal{K}_{\rm drag}=r\,\mathcal{K}_{\rm 5th}$,
where we find $r=-0.540\pm0.020$, significantly different from the matter-era estimate $r\simeq-0.66$. \\

Both potentials give essentially the same growth-equation coefficients: $c_5=0.657$, with $r=-0.540$ (exponential) and $r=-0.555$ (inverse power law). This is unsurprising, since neither the fifth-force term nor the drag depends on $V$ except through the background evolution. By contrast, the background amplitude does depend on the potential and its slope. At fixed cosmology, varying $\lambda$ yields $\mathcal{K}_{\rm bg}\propto \lambda^{-0.107\pm0.007}$ for the exponential potential, but $\mathcal{K}_{\rm bg}\propto \lambda^{+0.018\pm0.016}$ for the inverse power law, consistent with no $\lambda$ dependence. Because $\lambda$ enters differently in $e^{-\lambda\phi/\sqrt{8\pi}}$ and $\phi^{-\lambda}$, the chains prefer widely separated values (4.93 versus 0.83), and the response differs in both magnitude and sign. We therefore fit the two cases separately. Restoring physical units, we obtain
\ben
\begin{aligned}
\sum m_\nu\big|_{\rm exp} &\simeq
\Big[5.05\,f_c^{0.92}\Om_m^{0.86}h^{1.73}\lambda^{-0.11}
+ 2.24\,f_c^{2}\Om_m h^{2}\Big]\al^{2},\\[2pt]
\sum m_\nu\big|_{\rm pl} &\simeq
\Big[3.27\,f_c^{0.86}\Om_m^{0.57}h^{1.78}
+ 2.17\,f_c^{2}\Om_m h^{2}\Big]\al^{2},
\end{aligned}
\label{12}
\een
in eV. In each bracket, the first term is the background contribution and the second comes from the growth equation; the latter combines the fifth force and the drag through $c_5(1+r)$. The second term agrees between the two potentials at the $\sim3\%$ level, while the first is potential-dependent, cleanly separating what is fixed by the growth dynamics from what must be fitted in the background. 

Equation~(\ref{12}) reproduces the target relation without bias, with an rms deviation of $1.21\%$ and a worst-case deviation of $2.43\%$ for the exponential potential, and $1.64\%$ and $4.12\%$ for the power law. When applied to the six chains it predicts $C=0.879$--$0.932$ versus the $0.858$--$0.922$ obtained from the data-sector profile, i.e. higher by $1.0\%$--$2.9\%$. These are different estimators: the first comes from a model-space condition on $f\sigma_8$, and the second from profiling over the growth likelihood, so agreement at this level is better than what the construction itself guarantees.

However, we have two limitations that should be stated clearly. The power-law calibration covers only $\lambda=0.45$--$0.83$, since the gauge condition admits no solution above $\lambda\simeq0.9$, so the lack of any $\lambda$ dependence is established over a shorter lever arm than for the exponential case. By contrast, the exponential fit quoted spans $\lambda=3.0$ to $6.2$.

Equation~(\ref{12}) includes a small set of fitted, dimensionless coefficients that would be absent in a purely index-based scaling argument. We view this as the unavoidable price of adding the two extra contributions. The $\al^2$ scaling, the $f_c^2$ dependence of the fifth-force term, and the ratio $r$ that sets the strength of the opposing drag are derived, and the first two are independently confirmed by measurement. The remaining freedom is confined to the overall normalization of each term and the background power-law indices, because the background contribution is fixed by a shooting condition with no closed-form solution. 

In our model a single coupling, $\alpha$, controls the CDM drag, the running CDM mass, and the scalar-mediated fifth force, with their relative amplitudes fixed by the theory. An increase in $\sum m_\nu$ has also been reported in coupled-quintessence scenarios with explicit energy--momentum transfer~\cite{BeltranJimenez2026}, so the sign of the neutrino-mass correlation alone does not uniquely diagnose the interaction. The decomposition in Table~\ref{TableIV} is therefore a more discriminating test: a fluid-level coupling can alter the growth history without the associated Poisson enhancement, and hence would not reproduce the observed split among the three contributions.\\

The growth-only likelihood exhibits isolated secondary minima in $\chi^2$ at two or three distinct values of $\al$ per chain; these features persist when we rescan $\al$ on a 60-node seed grid and are therefore physical rather than numerical artifacts. To obtain a robust estimate of the growth-sector degeneracy slope we therefore use the Theil--Sen estimator, which is insensitive to such outliers~\cite{Theil1950,Sen1968}. The measured slope $C_{\rm meas}$ also depends on the $\al$-range used in the fit, because the marginalised $\sum m_\nu$--$\al^2$ relation is curved (Fig.~\ref{Fig6}): the local slope increases from $\simeq0.2$ near $\al^2=0.04$ to $\simeq1.2$ near $\al^2=0.30$, so any linear fit is necessarily a range-averaged approximation. For $\al^2\gtrsim0.32$ the posteriors become sparse and the local slope is no longer well constrained. Consistently, fitting only $\al^2>0.05$ yields a slope of $0.997$--$1.018$, while restricting to $\al^2>0.10$ gives $1.043$--$1.071$ (Table~\ref{TableIV}). Finally, this behavior is not driven by the oscillation-imposed lower bound $\sum m_\nu\ge 0.06\,{\rm eV}$: removing the $0.10\%$--$0.34\%$ of samples that sit on the floor shifts the full-range slope by less than $0.1\%$~\cite{NuFIT2024}.

The offset between the growth-only curve and Eq.~(\ref{12}) is likewise not an artifact of the oscillation floor, which is never saturated in this sector. In the scan we evaluate thirteen values of $\al$ spanning the $10$th--$90$th percentile range of its posterior mean, and for each value $\sum m_\nu$ minimises in the interior of parameter space rather than at the lower bound; the smallest best-fit value is $0.122\,{\rm eV}$. Even when the scan is extended to weaker coupling, the growth likelihood still prefers $\sum m_\nu=0.078\,{\rm eV}$ at $\al=0.05$, i.e. $\sim30\%$ above the floor (Fig.~\ref{Fig6}b). This residual offset is the neutrino mass favored by CMB lensing and RSD even when the coupling is switched off, and it is consistent with the low $\sigma_8\Omega_m^{0.25}$ preferred by ACT DR6~\cite{actdr6}.

In Table~\ref{TableIV} we estimate the coupling--neutrino-mass degeneracy slope $C_{\rm growth}$ using the Theil--Sen estimator over the $10$th--$90$th percentile range of $\al$, while holding all other parameters fixed at their posterior medians~\cite{Theil1950,Sen1968}. To isolate the physical origin of the trend we also report the changes in the slope obtained by switching off, in turn, the scalar-mediated fifth force and the coupling-induced drag. We label the resulting components as a background term, $C_{\rm bg}$, and perturbative increments, $C_{\rm 5th}$ and $C_{\rm drag}$. Their sum differs slightly from $C_{\rm growth}$ because the full profile involves a nonlinear minimisation. Finally, the analytic prediction $C_{\rm rel}$ from Eq.~(\ref{12}) matches the measured slopes to within $1\%$--$2.9\%$.

Figure~\ref{Fig6} illustrates the coupling--neutrino-mass degeneracy. In panel~(a) we compare the measured trend with the full analytic relation, and show that the growth-only contribution is substantially shallower, indicating that the background response dominates the total slope. The shaded band marks a $\pm10\%$ envelope around the analytic relation, whose residual relative to the model-space prediction is only $1.2\%$--$1.6\%$. In panel~(b) we adopt a two-sided prior, $\al\in[-1,1]$, and find that the degeneracy is approximately quadratic in $\al$, consistent with the expected $\al^2$ scaling of the fifth-force, drag, and background contributions. The curves are anchored at the fitted posterior intercepts and are shown only above the oscillation floor, $\sum m_\nu=0.06\,{\rm eV}$. 

\begin{table}[t]
\caption{\label{TableIV}Decomposition of the coupling-neutrino mass degeneracy slope $C\equiv\partial\sum m_\nu/\partial\alpha^2$.}
\begin{ruledtabular}
\begin{tabular}{llccccccc}
Data & $V$ & $f_c$ & $C_{\rm growth}$ & $C_{\rm bg}$ & $C_{\rm 5th}$ & $C_{\rm drag}$ & bg (\%) & $C_{\rm rel}$ (dev.\%) \\
\hline
DES-SN5YR & exp & $0.8275$ & $0.8666$ & $0.6550$ & $0.4486$ & $-0.2762$ & $75.6$ & $0.8900$ $(+2.7)$ \\
DES-SN5YR & pl & $0.8281$ & $0.9058$ & $0.6956$ & $0.4559$ & $-0.2800$ & $76.8$ & $0.9323$ $(+2.9)$ \\
Pantheon$+$ & exp & $0.8267$ & $0.8651$ & $0.6501$ & $0.4522$ & $-0.2744$ & $75.1$ & $0.8825$ $(+2.0)$ \\
Pantheon$+$ & pl & $0.8274$ & $0.9221$ & $0.7104$ & $0.4532$ & $-0.2825$ & $77.0$ & $0.9315$ $(+1.0)$ \\
Union3 & exp & $0.8267$ & $0.8584$ & $0.6419$ & $0.4477$ & $-0.2707$ & $74.8$ & $0.8789$ $(+2.4)$ \\
Union3 & pl & $0.8278$ & $0.9199$ & $0.7137$ & $0.4429$ & $-0.2821$ & $77.6$ & $0.9295$ $(+1.0)$ \\
\end{tabular}
\end{ruledtabular}
\end{table}

\subsection{Cosmological tensions}

\begin{figure}
    \centering
    \includegraphics[width=0.98\textwidth]{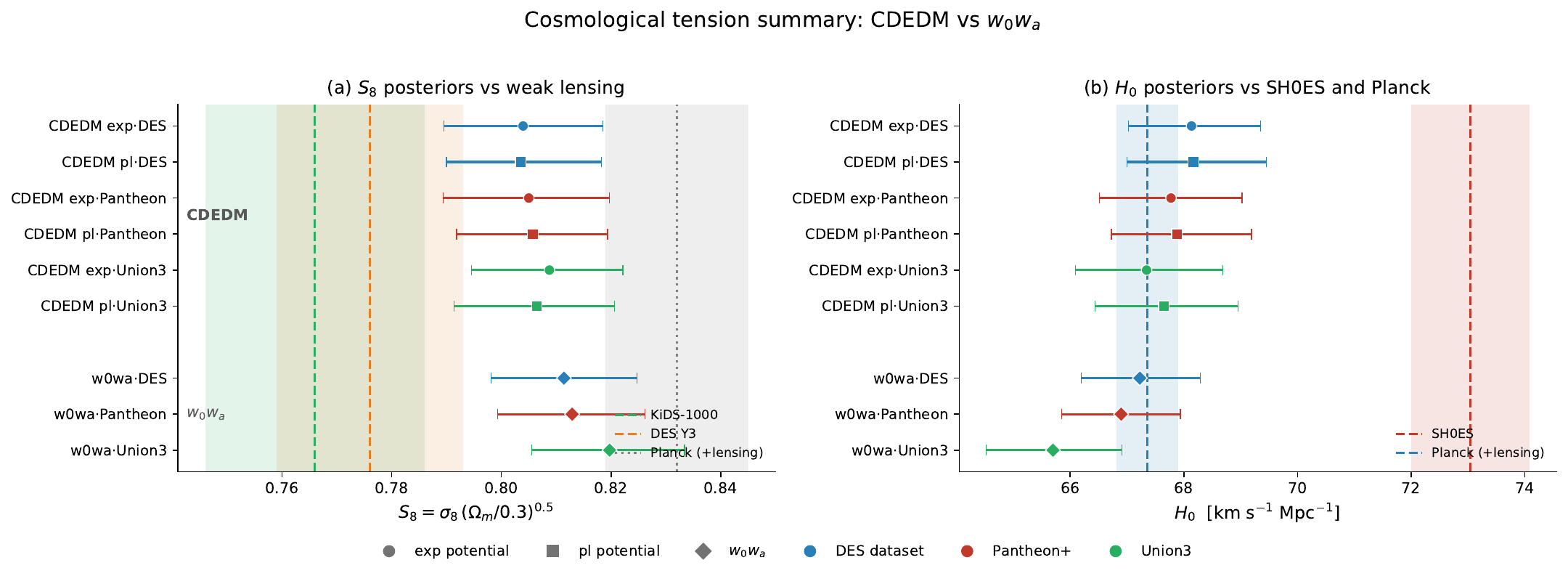}
    \caption{Marginalized $1\sigma$ constraints on $S_8$ (left) and $H_0$ (right) for CDEDM with exponential (circles) and inverse-power-law (squares) potentials, compared with $w_0w_a$CDM (diamonds). Results are shown for DES-SN5YR (blue), Pantheon$+$ (red), and Union3 (green), with the shaded bands denoting the corresponding reference constraints.}
    \label{Fig8}
\end{figure}

Figure~\ref{Fig8} shows the marginalized $1\sigma$ constraints on $H_0$ and $S_8$ in CDEDM (for both scalar potentials and all three SN compilations) compared with the corresponding $w_0w_a$CDM fits. For $H_0$, the CDEDM posteriors remain clustered around the \textit{Planck} value~\cite{Planck2020} and well below the SH0ES distance-ladder determination~\cite{Riess2022}, so the dark-sector interaction does not alleviate the Hubble tension~\cite{Verde2019, DiValentino2021}. This is unsurprising because we do not impose any SH0ES-based calibration. For $S_8$, CDEDM, likewise, does not fully reconcile CMB and weak-lensing results, but it does tend to shift the inferred values downward, toward the region favored by KiDS-1000~\cite{KiDS2x3pt} and DES~Y3~\cite{DES3x2pt}, instead of staying as close to the higher \textit{Planck}+lensing preference as some of the $w_0w_a$CDM fits. We therefore do not interpret CDEDM as a simultaneous solution to both tensions; rather, its main appeal is that it provides a theoretically motivated, ghost-free realization of phantom crossing while modifying the growth history in a way that can move $S_8$ in the weak-lensing direction.

\section{Conclusion}
\label{sec:conclusion}

We have investigated whether a phantom-like crossing of the effective dark-energy EoS can originate from a fundamental phantom degree of freedom. In the CDEDM model studied here, the dark-energy scalar field is strictly canonical, its kinetic term has the usual positive-definite sign, so the field carries no negative kinetic energy, and the apparent crossing is instead induced by energy exchange between dark matter and dark energy. At the level of the homogeneous background, this interaction-driven evolution is therefore degenerate with a purely phenomenological, time-varying dark-energy model. The degeneracy is broken only once perturbations and the recombination-era dynamics are included.

The most distinctive signal is the strong correlation we find between the dark-sector coupling and the summed neutrino mass. The scalar-mediated fifth force tends to boost the growth of structure, whereas neutrino free-streaming damps it, leading to a characteristic compensation between the two effects. Capturing this requires treating cold dark matter and baryons as separate species: only CDM feels the fifth force, and the resulting CDM--baryon offset modifies how the total matter field responds. In this way, the coupling--$\sum m_\nu$ degeneracy acquires a clear microscopic origin, and it is recovered robustly across all three supernova compilations and for both choices of scalar potential. When the coupling is allowed to take either sign, we also recover the expected approximately quadratic dependence on its magnitude.

A key result is that the fifth force alone does not set the measured coupling-neutrino-mass degeneracy. Instead, the coupling influences structure growth through three linked contributions: the modified background evolution, the fifth force, and the coupling-induced drag. We find that the background response accounts for roughly three quarters of the inferred slope, while the drag partially offsets the enhancement produced by the fifth force. The fact that this decomposition remains stable across the different supernova compilations and scalar potentials indicates that the degeneracy is a robust consequence of the coupled dynamics, not an artifact of any single data set.

The interaction also leaves a clear early-time imprint. Because the dark-matter mass evolves with the scalar field, the dark-matter density at recombination can differ significantly from its value today. The CMB acoustic scale and sound horizon must therefore be computed using the evolving dark-matter density. Approximating the dark-matter density as constant can bias the inferred coupling, even though the coupled model remains consistent with the acoustic-scale constraint.

Taken together, these findings reinforce that an effective phantom crossing is not, by itself, evidence for a fundamental phantom field. In the interacting scenario, the crossing is accompanied by additional, testable information encoded in the growth of structure and in the evolution of the dark-matter density toward recombination. In particular, the coupling-neutrino-mass degeneracy, together with its decomposition into background, fifth-force, and drag components, provides a more specific fingerprint of the underlying dark-sector interaction than the expansion history alone and offers a practical route to distinguish a microscopic interaction from a purely kinematic parametrization of dark-energy evolution.

Our results therefore show that an apparent phantom-like behavior can still encode information about the microscopic origin of cosmic acceleration, even when the background expansion history (distance measurements) is unable to discriminate between models. Upcoming measurements of structure growth, via improved redshift-space distortions, weak lensing, and CMB lensing, together with high-precision distance probes (e.g. stage-IV surveys such as \textit{Euclid}~\cite{Euclid2012}) can be able to test the predicted $\sum m_\nu\propto f_c^2\alpha^2$ scaling and, in doing so, help distinguish an interacting dark sector from a purely kinematic phantom-crossing description; however, a detailed forecast is beyond the scope of the present work.

\begin{acknowledgments}
G.M. (IUCAA Associate) acknowledges the Inter-University Centre for Astronomy and Astrophysics (IUCAA), Pune, India, for facilitating part of this work during his visit. G.M. thanks the COST Association (CA21136 CosmoVerse), European Union, for the opportunity to participate in this international association as a group member. G.M. expresses his sincere gratitude to all undergraduate, postgraduate, and doctoral students, as well as to his teachers, collaborators, and well-wishers, whose support has greatly enriched his academic journey. Beyond academia, G.M. is deeply grateful to a special presence whose understanding, encouragement, and companionship have quietly brought meaning and strength to this journey, and who helped him through a particularly difficult period in his life.
S.G. is deeply grateful to G.M. for his continued support and encouragement in his journey as a researcher.
 \\

\textbf{Conflicts of interest:} The authors declare no conflicts of interest.\\

\textbf{Funding information:} Not available.\\

\textbf{Data availability:} The data used in this study are readily accessible from public sources for validation of our model; however, we did not generate any new data sets for this research.\\

\end{acknowledgments}

\appendix



\section{Shooting for the gauge condition}
\label{sec:shooting}

We fix the field normalization by setting $\phi_0\equiv 1\,M_{\rm pl}$, which removes the degeneracy between $\lambda$ (or $\alpha$) and $\phi_0$. For each parameter vector we then solve for the initial value $\phi_{\rm ini}$ such that $\phi(N=0)=1$. We first evaluate 40 logarithmically spaced trial values in $\phi_{\rm ini}\in[0.02,8]$ to locate a sign-changing bracket, and then refine the root with 24 bisection steps, corresponding to a resolution of $\sim 6\times 10^{-8}$. This initial scan is necessary because, after evolving the coupling over the full integration range, the root can lie on either side of $\phi_0=1$; using a fixed bracket can therefore miss the root and instead return one of the bracket endpoints.

We set the initial velocity to its quasi-static attractor value,
\ben
v_{\rm QS} = -\frac{V_{,\phi} + 3H_0^2\Om_{c0}(1+z_{\rm ini})^3 f_{,\phi}/f_0}{3H(z_{\rm ini})},
\label{A5}
\een
which is accurate deep in the matter-dominated regime. Parameter samples for which no such root exists are assigned $\ln\La=-\infty$. We evolve the background using the fifth-order Tsitouras Runge--Kutta method (Tsit5)~\cite{Tsitouras2011} with ${\rm rtol}=10^{-7}$ and ${\rm atol}=10^{-9}$, allowing up to $3\times10^4$ integration steps. All calculations are performed in 64-bit precision with just-in-time~(JIT) compilation.

\section{Determination of $V_0$}
\label{sec:V0}

We do not sample $V_0$. Instead, for each parameter vector we determine it via a damped fixed-point iteration coupled to the shooting solve. Denoting by $\Om_{\rm de}^{\rm target}$ the dark-energy density required by exact flatness, we write
\ben
V_0^{(i+1)} = 0.8\,V_0^{(i)} + 0.2\Big[3H_0^2\Om_{\rm de}^{\rm target}
              - \tfrac12\big(\dot\phi_0^{(i)}\big)^2\Big].
\label{A6}
\een
We iterate until the flatness residual drops below $10^{-5}$, or until a maximum of 30 iterations is reached. Because the bracketing interval for the shooting solve can become invalid as $V_0$ changes, we repeat the bracketing scan and bisection whenever the first pass leaves a residual above $10^{-4}$. This second pass prevents isolated gaps in otherwise allowed regions of parameter space caused by an outdated bracket.

We reject a sample if the fixed-point iteration fails, if $V_0\le10^{-8}$, if the flatness residual exceeds $10^{-4}$, or if $|\phi(N{=}0)-1|>10^{-3}$. With these criteria, the best-fit points have a flatness residual of $\sim2\times10^{-5}$ and a shooting residual of $\sim4\times10^{-9}$.

This procedure sets the upper bound on $\lam$ for the inverse-power-law model. Among 400 cosmologies drawn from the remaining prior volume, the fraction admitting a flatness fixed point decreases from $31\%$ at $\lam=1.20$ to $9.5\%$, $2.3\%$, and $0.75\%$ at $\lam=1.30$, $1.40$, and $1.45$, respectively, and vanishes for $\lam\ge1.50$. Thus, the upper prior edge is set by the model's existence conditions rather than by the data. 

\section{Posterior distributions of the reference models}\label{sec:reference}

Both reference models ($w_0w_a$CDM and $\Lam$CDM) are fitted to the same data combination as CDEDM, namely the CMB distance priors, DESI~DR2 BAO, the RSD compilation, the ACT lensing constraint, and one of the three supernova samples. The corresponding marginal posteriors are listed in Table~\ref{TableV}. Since we sample $\sum m_\nu$ and $N_{\rm eff}$ in both cases, the parameter counts are five for $\Lam$CDM and seven for $w_0w_a$CDM, with the latter matching CDEDM.

The $\Lam$CDM constraints are essentially insensitive to the choice of supernova compilation. Across DES-SN5YR, Pantheon$+$, and Union3, $\Om_{m0}$ varies only from $0.3011$ to $0.3021$ and $H_0$ from $68.63$ to $68.79\,$km\,s$^{-1}$Mpc$^{-1}$; the remaining sampled parameters are likewise stable, with $\om_b\simeq0.02240$, $N_{\rm eff}\simeq3.17$, and a nearly identical $95\%$ bound $\sum m_\nu\lesssim0.15\,$eV. This behavior is expected: with $w=-1$ the distance-redshift relation is fixed in shape, and the BAO+CMB data already determine $\Om_{m0}$ and $\om_b$ more tightly constrained than the supernovae can. The supernova likelihood then mainly constrains its calibration nuisance, so swapping compilations has negligible impact. Any stronger dataset dependence seen in extended models therefore reflects genuine background dynamics rather than an artifact of the supernova data choice.

The $w_0w_a$CDM posterior behaves very differently. For all three supernova compilations the data prefer $w_0>-1$ and $w_a<0$, i.e. an equation of state that is non-phantom today but evolves to more negative values at higher redshift and therefore crosses $w=-1$ at low $z$. Concretely, we obtain $w_0=-0.815^{+0.061}_{-0.059}$, $-0.784^{+0.058}_{-0.057}$, and $-0.664^{+0.092}_{-0.091}$ with $w_a=-0.697^{+0.253}_{-0.275}$, $-0.767^{+0.251}_{-0.256}$, and $-1.124^{+0.337}_{-0.340}$ for DES-SN5YR, Pantheon$+$, and Union3, respectively (Table~\ref{TableV}). The posterior mass in this quadrant is $P(w_0>-1,\,w_a<0)=0.998$, $1.000$, and $1.000$, so the preference for a crossing is strong. The corresponding crossing redshifts are $z_\times=0.36^{+0.10}_{-0.07}$, $0.39^{+0.12}_{-0.07}$, and $0.43^{+0.09}_{-0.06}$, systematically earlier than in CDEDM, as also illustrated in Fig.~\ref{Fig2}. Among the three, Union3 drives the largest departure from $\Lam$CDM, consistent with its higher-redshift reach and larger statistical uncertainties.

Relative to $\Lam$CDM, the $w_0w_a$ extension lowers the inferred expansion rate: $H_0$ shifts from $\simeq 68.7$ to $67.23^{+1.07}_{-1.03}$, $66.90^{+1.04}_{-1.05}$, and $65.70^{+1.21}_{-1.17}$ for DES-SN5YR, Pantheon$+$, and Union3, respectively (Table~\ref{TableV}). The kinematic freedom therefore moves the inference away from local distance-ladder determinations rather than toward them. At the same time, it increases the clustering amplitude, with $S_8$ rising from $0.801$ in $\Lam$CDM to $0.8114^{+0.0133}_{-0.0133}$, $0.8129^{+0.0132}_{-0.0136}$, and $0.8197^{+0.0137}_{-0.0142}$, i.e. further from typical weak-lensing preferences. Overall, a purely kinematic modification improves the fit to the distance data (Table~\ref{TableII}) while worsening both the $H_0$ and $S_8$ tensions.

Neutrino masses further separate the models. In the reference fits the $95\%$ bounds are $\sum m_\nu<0.155$, $0.155$, and $0.152\,$eV in $\Lam$CDM and $\sum m_\nu<0.189$, $0.189$, and $0.190\,$eV in $w_0w_a$CDM (Table~\ref{TableV}), to be compared with $0.459$--$0.508\,$eV in CDEDM (Table~\ref{TableI}). The coupled model therefore relaxes the inferred upper bound by a factor of ${\sim}3$, which reflects the $\alpha$--$\sum m_\nu$ degeneracy discussed in Sec.~\ref{sec:degeneracy}: increasing $\alpha$ enhances growth and can be compensated by a larger neutrino mass. Finally, $N_{\rm eff}$ remains consistent with the standard value in all cases, spanning $3.040$--$3.182$ with typical uncertainties of ${\sim}0.19$ (Table~\ref{TableV}). 

\begin{figure*}[t]
\centering
\includegraphics[width=0.49\textwidth]{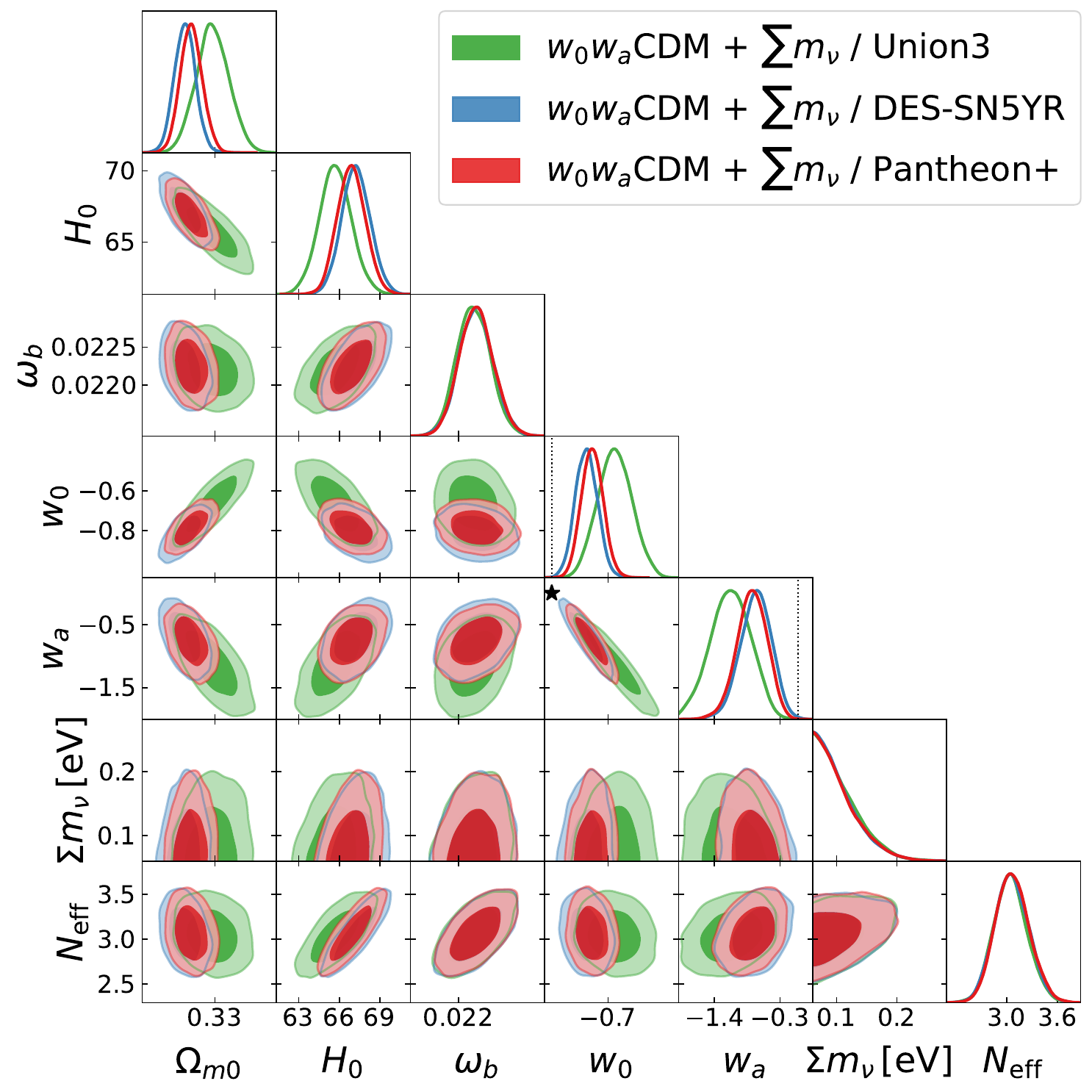}\hfill
\includegraphics[width=0.49\textwidth]{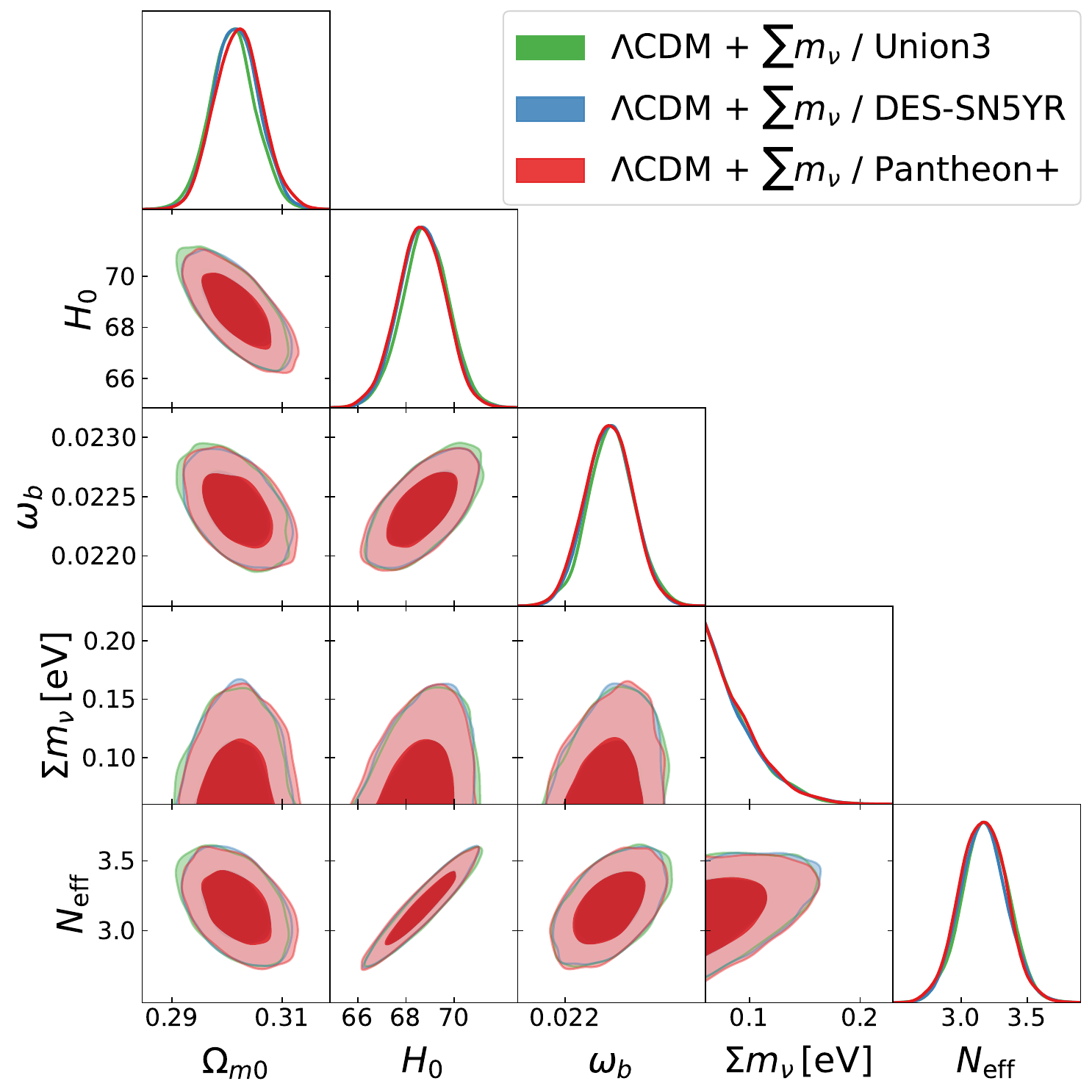}
\caption{\label{Fig7}Posteriors for the reference models, for the three
supernova compilations. \textit{Left:} $w_0w_a$CDM, with the star marking
$(w_0,w_a)=(-1,0)$. \textit{Right:} $\Lam$CDM, whose contours coincide to
within $0.24\sigma$. Contours enclose $68\%$ and $95\%$.}
\end{figure*}

\begin{table}[t]
\caption{\label{TableV}Marginalised constraints on the two reference models, for the three supernova compilations along with CMB distance-prior, RSD data, DESI DR2 BAO data and ACT lensing constraint. Uncertainties are $68\%$ credible intervals about the median, except $\sum m_\nu$ which is quoted as a $95\%$ interval.}
\centering
\setlength{\tabcolsep}{2pt}
\scriptsize
\begin{ruledtabular}
\resizebox{\linewidth}{!}{%
\begin{tabular}{lccccccc}
Data & $\Om_{m0}$ & $H_0$ & $\om_b$ & $N_{\rm eff}$
 & $\sum m_\nu$~[eV] & $w_0$ & $w_a$ \\
 & & [km\,s$^{-1}$Mpc$^{-1}$] & & & ($95\%$ C.I.) & & \\
\hline
\multicolumn{8}{l}{$w_0w_a$CDM} \\
DES-SN5YR & $0.3138^{+0.0055}_{-0.0057}$ & $67.23^{+1.07}_{-1.03}$ & $0.02224^{+0.00024}_{-0.00023}$ & $3.053^{+0.209}_{-0.197}$ & $[0.062,\,0.189]$ & $-0.815^{+0.061}_{-0.059}$ & $-0.697^{+0.253}_{-0.275}$ \\
Pantheon$+$ & $0.3171^{+0.0060}_{-0.0058}$ & $66.90^{+1.04}_{-1.05}$ & $0.02225^{+0.00024}_{-0.00024}$ & $3.060^{+0.205}_{-0.192}$ & $[0.062,\,0.189]$ & $-0.784^{+0.058}_{-0.057}$ & $-0.767^{+0.251}_{-0.256}$ \\
Union3 & $0.3286^{+0.0090}_{-0.0088}$ & $65.70^{+1.21}_{-1.17}$ & $0.02221^{+0.00023}_{-0.00023}$ & $3.040^{+0.199}_{-0.194}$ & $[0.062,\,0.190]$ & $-0.664^{+0.092}_{-0.091}$ & $-1.124^{+0.337}_{-0.340}$ \\
\hline
\multicolumn{8}{l}{$\Lam$CDM} \\
DES-SN5YR & $0.3016^{+0.0041}_{-0.0040}$ & $68.69^{+0.96}_{-0.96}$ & $0.02240^{+0.00021}_{-0.00021}$ & $3.168^{+0.174}_{-0.175}$ & $[0.061,\,0.155]$ & $-1$ & $0$ \\
Pantheon$+$ & $0.3021^{+0.0042}_{-0.0043}$ & $68.63^{+0.99}_{-0.98}$ & $0.02239^{+0.00021}_{-0.00022}$ & $3.164^{+0.175}_{-0.177}$ & $[0.061,\,0.155]$ & $-1$ & $0$ \\
Union3 & $0.3011^{+0.0041}_{-0.0040}$ & $68.79^{+0.98}_{-0.95}$ & $0.02241^{+0.00021}_{-0.00020}$ & $3.182^{+0.178}_{-0.171}$ & $[0.061,\,0.152]$ & $-1$ & $0$ \\
\end{tabular}%
}
\end{ruledtabular}
\end{table}

\section{Posterior predictive checks}
\label{sec:ppc}

\begin{figure}
    \centering
    \includegraphics[width=0.70\textwidth]{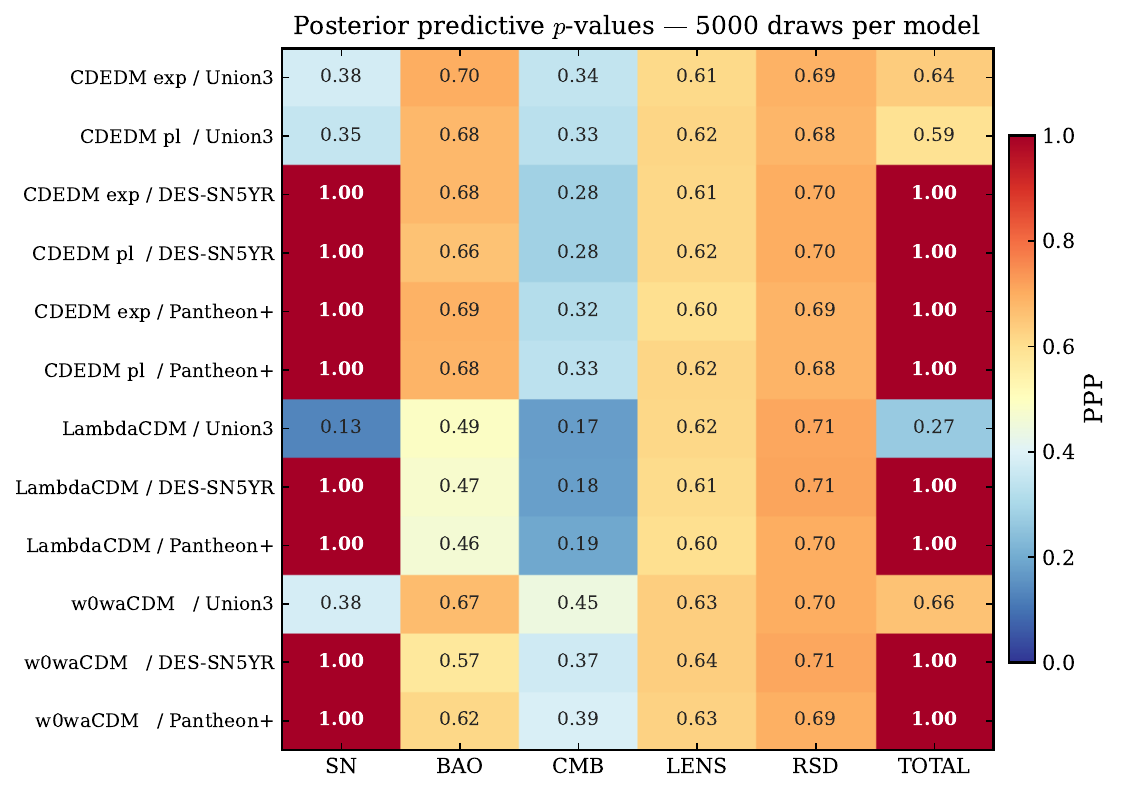}
    \caption{posterior predictive check across all model and dataset compilation}
    \label{FigS5}
\end{figure}

As an additional check on the adequacy of the fitted models, we perform posterior predictive checks using 5000 draws from the posterior predictive distribution for each model-dataset combination. The resulting posterior predictive $p$-values are shown in Fig.~\ref{FigS5}. For most observables, these values lie well away from the extreme tails, indicating no clear evidence that the observed data are inconsistent with the corresponding posterior predictive distributions. Our CDEDM models yield broadly similar $p$-values for the two choices of scalar potential and across the three supernova compilations, suggesting that our conclusions are not driven by an anomalous fit to any single dataset. We also find that the CDEDM and $w_0w_a$CDM models exhibit comparable posterior predictive behaviour for the common observables, consistent with our background-degeneracy results. Some $p$-values are close to unity, particularly for the supernova and total statistics, but this should not be interpreted as evidence of a perfect fit; rather, it indicates that the observed discrepancy statistic is smaller than that obtained in most posterior predictive realizations. Overall, these checks show no systematic evidence for a severe model-data mismatch and provide an independent consistency test of the likelihood analysis.

\section{The sign of the coupling}
\label{sec:twosided}
In our original analysis, we impose $\al\in[0,1]$. This restriction is physically motivated: achieving $w_{\rm eff}<-1$ requires $\al>0$, so the negative branch does not realise the phantom-crossing mechanism. However, it comes with a statistical cost. When the null hypothesis lies on the prior boundary, a displacement in units of $\sigma$ is not well-defined, and the Savage--Dickey density ratio is only an approximation to the nested-model Bayes factor. Two questions therefore remain, neither of which can be answered within the one-sided runs: does truncating the prior at $\al=0$ artificially induce a preference for the coupled model, and is $\al=0$ actually disfavoured by the data?

We therefore repeated all six analyses with a two-sided prior, $\al\sim U[-1,1]$, keeping everything else unchanged.

\begin{table}
\caption{\label{TableA2}Model comparison under the two-sided coupling prior $\al\sim U[-1,1]$. Here, $\Del$DIC is quoted against $\Lam$CDM.}
\begin{ruledtabular}
\begin{tabular}{llccccc}
Model & Data & $k$ & $\chi^2_\nu$ & $\Del\ln Z_{w_0w_a}$ & $\Del\ln Z_{\Lam}$ & $\Del$DIC \\
\hline
$w_0w_a$ & DES  & 7 & 0.8962 & $0$    & $-0.05$ &  $-6.3$ \\
exp      & DES  & 7 & 0.8963 & $2.21$ & $2.16$  &  $-6.3$ \\
pl       & DES  & 7 & 0.8963 & $2.07$ & $2.01$  &  $-6.4$ \\
\hline
$w_0w_a$ & Pan. & 7 & 0.9023 & $0$    & $2.28$  &  $-10.5$ \\
exp      & Pan. & 7 & 0.9022 & $1.95$ & $4.23$  &  $-10.9$ \\
pl       & Pan. & 7 & 0.9022 & $1.68$ & $3.96$  &  $-11.2$ \\
\hline
$w_0w_a$ & Un3  & 7 & 0.9113 & $0$    & $2.22$  &  $-10.3$ \\
exp      & Un3  & 7 & 0.9292 & $1.31$ & $3.53$  &  $-9.3$ \\
pl       & Un3  & 7 & 0.9353 & $0.38$ & $2.60$  &  $-8.3$ \\
\end{tabular}
\end{ruledtabular}
\end{table}

Table~\ref{TableA2} repeats Table~\ref{TableI} with the wider prior. Even so, every coupled model remains favoured over $w_0w_a$CDM at equal parameter count. The only notable exception is the power-law model for Union3, which drops to $\Del\ln Z_{w_0w_a}=0.38$ and should be regarded as inconclusive rather than supportive.

The uniform reduction of $0.36$--$0.63$ relative to Table~\ref{TableI} does not represent weaker evidence; it is the expected prior-volume penalty from doubling the sampled range. Restricting the two-sided evidence back to the positive branch,
\be
\ln Z_{+} = \ln Z_{\rm tot} + \ln w_{+} - \ln P(\al>0\,|\,\pi),
\label{A13}
\ee
with $P(\al>0|\pi)=1/2$, reproduces the one-sided values of Table~\ref{TableI} to within $+0.03$, $+0.16$, $+0.08$ (exponential) and $+0.23$, $+0.32$, $+0.20$ (power law) for DES-SN5YR, Pantheon$+$, and Union3. These are independent runs with $\sigma_{\ln Z}\simeq 0.09$--$0.13$, so the agreement is within $0.2$--$2.0\sigma$ throughout. We therefore conclude that truncating the prior at $\al=0$ did not manufacture the preference.

\begin{table}
\caption{\label{TableA4}Coupling constraints with the null interior to the prior. $w_+$ is defined as the fraction of posterior mass at $\al>0$.}
\begin{ruledtabular}
\begin{tabular}{llcccc}
$V(\phi)$ & Data & $\al$ (median) & $95\%$ interval & $w_+$ & $\mathrm{BF}_{10}$ \\
\hline
exp & DES  & $0.417$ & $[-0.030,\,0.642]$ & $0.971$ & $3.8$ \\
exp & Pan. & $0.444$ & $[\phantom{-}0.087,\,0.662]$ & $0.982$ & $7.5$ \\
exp & Un3  & $0.481$ & $[\phantom{-}0.202,\,0.695]$ & $0.994$ & $14.2$ \\
\hline
pl  & DES  & $0.412$ & $[-0.012,\,0.635]$ & $0.974$ & $5.2$ \\
pl  & Pan. & $0.421$ & $[\phantom{-}0.096,\,0.641]$ & $0.985$ & $8.4$ \\
pl  & Un3  & $0.432$ & $[\phantom{-}0.138,\,0.633]$ & $0.987$ & $7.9$ \\
\end{tabular}
\end{ruledtabular}
\end{table}

With $\al=0$ now interior, the Savage--Dickey ratio yields a well-posed nested-model Bayes factor. Table~\ref{TableA4} gives $\mathrm{BF}_{10}=3.8$--$14.2$, which is substantial to strong on the Jeffreys scale~\cite{Jeffrey1961}, with $97.1$--$99.4\%$ of the posterior mass at $\al>0$. The $95\%$ intervals exclude zero for four of the six combinations, although the two DES-SN5YR cases include it marginally.

\end{document}